\RequirePackage{fix-cm}
\documentclass[smallextended]{svjour3}       
\pdfoutput=1
\usepackage[pdftex]{graphicx}
\DeclareGraphicsExtensions{.pdf,.jpeg,.jpg,.png}
\usepackage{subfigure}
\usepackage[latin1]{inputenc}
\usepackage{lmodern}
\usepackage{textcomp}
\usepackage[version=3]{mhchem}
\journalname{Experimental Astronomy}
\begin{document}

\title{Soft Proton Scattering Efficiency Measurements on X-Ray Mirror Shells
}


\author{Sebastian Diebold \and
				Chris Tenzer \and
				Emanuele Perinati \and
				Andrea Santangelo \and
				Michael Freyberg \and
				Peter Friedrich \and
				Josef Jochum
}


\institute{S. Diebold, E. Perinati, A. Santangelo, C. Tenzer \at
              Institute for Astronomy and Astrophysics,\\Kepler Center for Astro and Particle Physics,\\University of T\"ubingen, Sand 1, 72076 T\"ubingen\\
              Tel.: +49-7071-29-78609\\
              Fax: +49-7071-29-3458\\
              \email{diebold@astro.uni-tuebingen.de}\\
           \and
           M. Freyberg, P. Friedrich \at
              Max Planck Institute for Extraterrestrial Physics,\\Giessenbachstrasse 1, 85748 Garching\\
           \and
           J. Jochum \at
              Physics Institute T\"ubingen,\\Kepler Center for Astro and Particle Physics,\\University of T\"ubingen, Auf der Morgenstelle 14, 72076 T\"ubingen\\
}

\date{Received: date / Accepted: date}

\maketitle

\begin{abstract}
In-orbit experience has shown that soft protons are funneled more efficiently through focusing Wolter-type optics of X-ray observatories than simulations predicted. These protons can degrade the performance of solid-state X-ray detectors and contribute to the instrumental background. Since laboratory measurements of the scattering process are rare, an experiment for grazing angles has been set up at the accelerator facility of the University of T\"ubingen. Systematic measurements at incidence angles ranging from 0.3\textdegree~to 1.2\textdegree~with proton energies around 250\,keV, 500\,keV, and 1\,MeV have been carried out. Parts of spare mirror shells of the \textit{eROSITA} (extended ROentgen Survey with an Imaging Telescope Array) instrument have been used as scattering targets. This publication comprises a detailed description of the setup, the calibration and normalization methods, and the scattering efficiency and energy loss results. A comparison of the results with a theoretical scattering description and with simulations is included as well.

\end{abstract}

\section{Introduction}\label{sec:intro}
Astronomical X-ray satellites are exposed to various kinds of particle radiation: trapped protons and electrons, solar wind particles, and energetic particles of extrasolar origin. Although the radiation environment is highly variable and depends strongly on the orbit, usually protons have the largest impact on the mission. The effects of proton radiation on the performance of X-ray instruments can be sub-divided into two classes: degradation of the detector performance and contribution to the background of astronomical observations~\textendash~either via direct interaction with the detector, or, more likely, via the production of secondary particles and fluorescence in the surroundings of the detector. Therefore, detailed studies of the expected proton flux at the detector location enable not only an assessment of possible radiation damages, moreover they are crucial to evaluate the scientific performance of an instrument, i.e. the signal-to-noise ratio.

Since irradiation tests of completely assembled X-ray instruments are hardly feasible, Monte Carlo based radiation transport codes are used to simulate the interactions of the orbital radiation with the spacecraft and the scientific instrumentation. Nowadays most simulation codes and toolkits are evolved to such an extent that uncertainties of the orbital radiation models tend to limit the accuracy. Nevertheless, the treatment of soft proton scattering at grazing incidence angles, as it occurs for Wolter-type focusing X-ray mirrors, is still a matter of discussion. The problem of soft proton funneling has been discovered shortly after the launch of the \textit{Chandra} X-ray observatory \cite{Chandra} in 1999, when a surprisingly rapid degradation of the front-illuminated CCDs (charge coupled devices) of the \textit{ACIS} instrument occurred \cite{Chandra_ACIS}. Measurements with \textit{XMM-Newton} \cite{XMM}, which was launched six months after \textit{Chandra}, showed a soft proton flux more than a factor of three larger than expected from simulations \cite{Kendziorra_2000}. In the following years, the scattering description for grazing angles by Firsov \cite{Firsov}, which is based on the surface plasmon model, has been implemented in the \textsc{Geant4} toolkit\footnote{http://geant4.cern.ch} \cite{Geant4} to solve the discrepancy \cite{Lei_2004}. Nevertheless, experimental data with X-ray mirrors for validation are still rare. So far only one experiment, performed at the accelerator facility of the Harvard University, has been measuring soft proton scattering on X-ray mirrors \cite{XMM-RGS,Chandra_XMM}, but it was carried out under enormous time pressure between the discovery of the \textit{Chandra} incident and the launch of \textit{XMM-Newton}. Therefore, it covered only the parameter space that was relevant to assess possible degradation of the \textit{XMM-Newton} instruments.

Since the scattering of soft protons from the orbital environment is an issue for future missions as well, the lack of experimental data has driven the construction of a new setup to measure grazing angle scattering efficiencies of soft protons over a wide range of parameters and, as a secondary goal, to determine the energy loss. It is based on a previous irradiation setup \cite{Diebold_2013} that has been used to study the effects of soft proton radiation on detector prototypes for the \textit{LOFT} (Large Observatory For x-ray Timing) space mission \cite{LOFT}. The intention of the scattering measurements is, on the one hand, to validate the application of the Firsov description to X-ray mirrors as well as the use of the \textsc{TRIM} code\footnote{Part of the \textsc{SRIM} package; http://www.srim.org} \cite{SRIM} for this purpose and, on the other hand, to generate input parameters for a ray tracing code currently under development to simulate the propagation of soft protons through X-ray telescopes \cite{Perinati_2014}. All results presented within this publication have been obtained with parts of spare mirror shells of the \textit{eROSITA} (extended Roentgen Survey with an Imaging Telescope Array) instrument \cite{eROSITA}. \textit{eROSITA} features seven Wolter type-I X-ray telescopes, each consisting of 54 nested mirror shells. It is the main instrument of the \textit{SRG} (Spectrum-Roentgen-Gamma) satellite, which is scheduled for a launch in 2016 and will be placed in an orbit around the Lagrangian point L2 of the Earth-Sun-system to perform an all-sky survey. Since it will be the first X-ray telescope operating at L2, where the (soft) proton flux has not been measured so far, it represents also a pathfinder for future X-ray missions at L2, e.g. \textit{ATHENA} (Advanced Telescope for High-ENergy Astrophysics) \cite{Athena}.

A comprehensive description of the scattering experiment and the calibration methods is given in Section~\ref{sec:setup}. In Section~\ref{sec:meas} the measurement procedure and parameters are explained, while the results are presented and discussed in Section~\ref{sec:results}. Section~\ref{sec:conc} concludes with a summary and an outlook on the application of the results as well as on future measurements.


\section{Experimental Setup}\label{sec:setup}

\subsection{Accelerator Facility}\label{ssec:accelerator}
The scattering setup for grazing angles has been implemented at the ion accelerator facility of the University of T\"ubingen. The accelerator is a 3\,MV single-ended Van de Graaff (HVEC KN-3000), capable of producing light ion beams in the energy range of 400\,keV\textendash2.5\,MeV. The beam energy is determined by measuring the terminal voltage with a generating voltmeter (GVM) \cite{IonBeamGVM}. The GVM itself is calibrated via measuring the yield of resonant nuclear reactions and threshold reactions while scanning the terminal voltage \cite{AccCalib}. The beam current is adjustable from a few nanoampere up to several tens of microampere.

The ions are produced in a radio frequency ion source within the high voltage terminal. An analyzing magnet bends the beam by 95\textdegree~to remove contaminations from ionized residual gas and partially ionized molecules via momentum separation. This analyzer is also used to stabilize the terminal voltage on the $10^{-3}$ level by feeding back the beam position after the 95\textdegree-bend to the voltage regulation system.

\begin{figure*}
  \includegraphics[width=\textwidth]{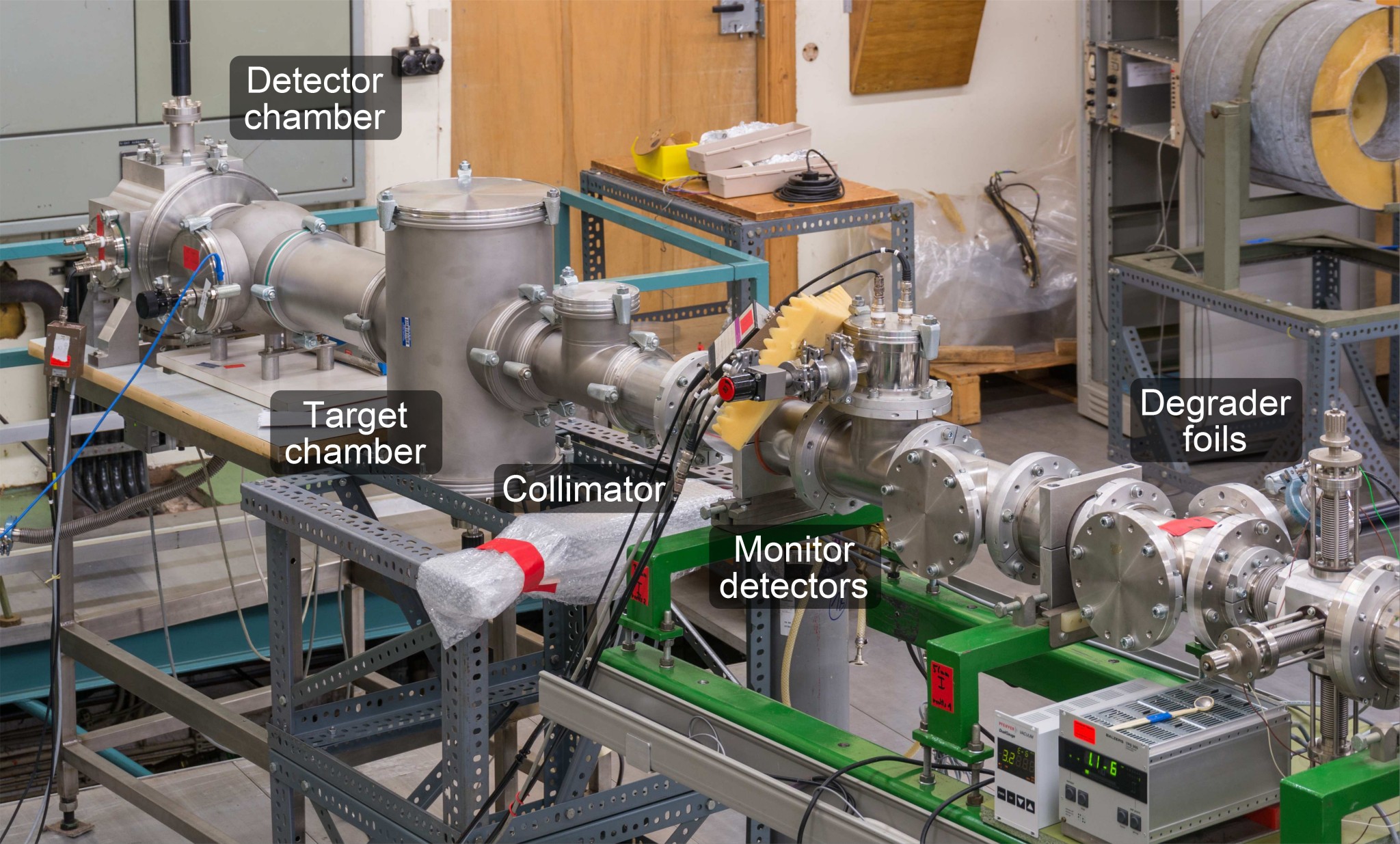}
	\caption{Overview of the scattering experiment at beam line 3 of the accelerator facility of the University of T\"ubingen. The beam direction is from the right hand side, where the last slit is visible, through the collimator in the center to the detector chamber on the far left. The scattering target is mounted in the chamber downstream of the collimator. Degrader foils widen the beam spatially to provide flux to two detectors at the collimator entry, which monitor the incident proton flux.}
	\label{fig:picture}
\end{figure*}

The facility features six beam lines, selectable via a switching magnet. Various dipole magnets are distributed along the beam lines to bend and shift the beam. Two pairs of double quadrupole magnets per beam line allow to refocus the beam. An overview of the small angle scattering setup installed at the end of beam line 3 is given in Fig.~\ref{fig:picture}.

\subsection{Beam line setup}\label{ssec:blsetup}
\begin{figure*}
  \includegraphics[width=\textwidth]{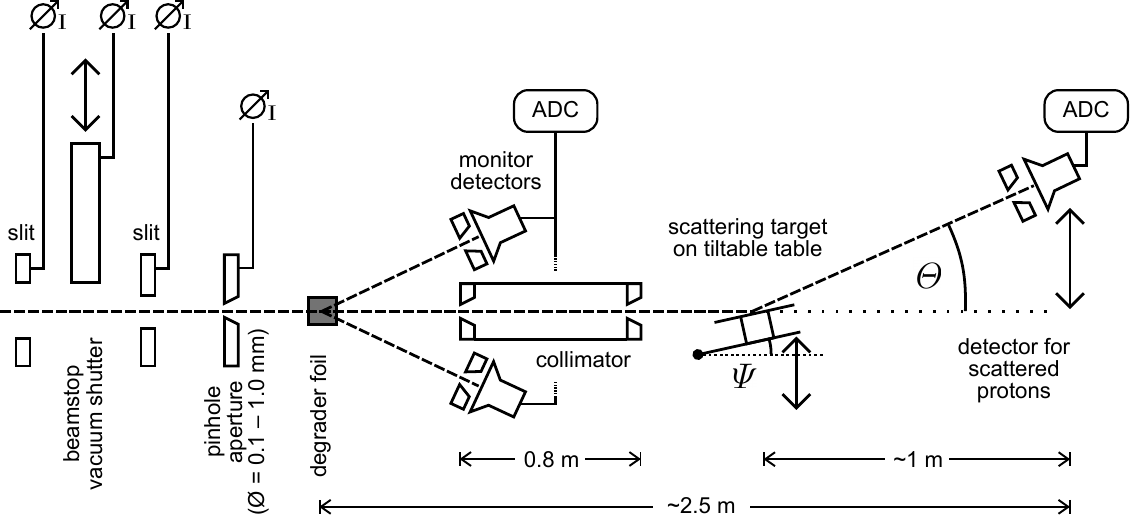}
	\caption{Schematic of the small angle scattering setup. Two pairs of slits are used to verify the alignment of the proton beam with the setup. A pinhole aperture reduces the beam current and ensures a well-defined beam spot on the metal degrader foil. The foil spatially widens the beam to provide flux to the monitor detectors. A collimator focuses a part of the widened beam on the target, which is mounted on a tiltable table to allow a change of the incidence angle $\Psi$. The scattering angle $\Theta$ is defined by the position of the detector at the end of the beam line.}
	\label{fig:scheme}
\end{figure*}

\begin{figure*}
  \includegraphics[width=\textwidth]{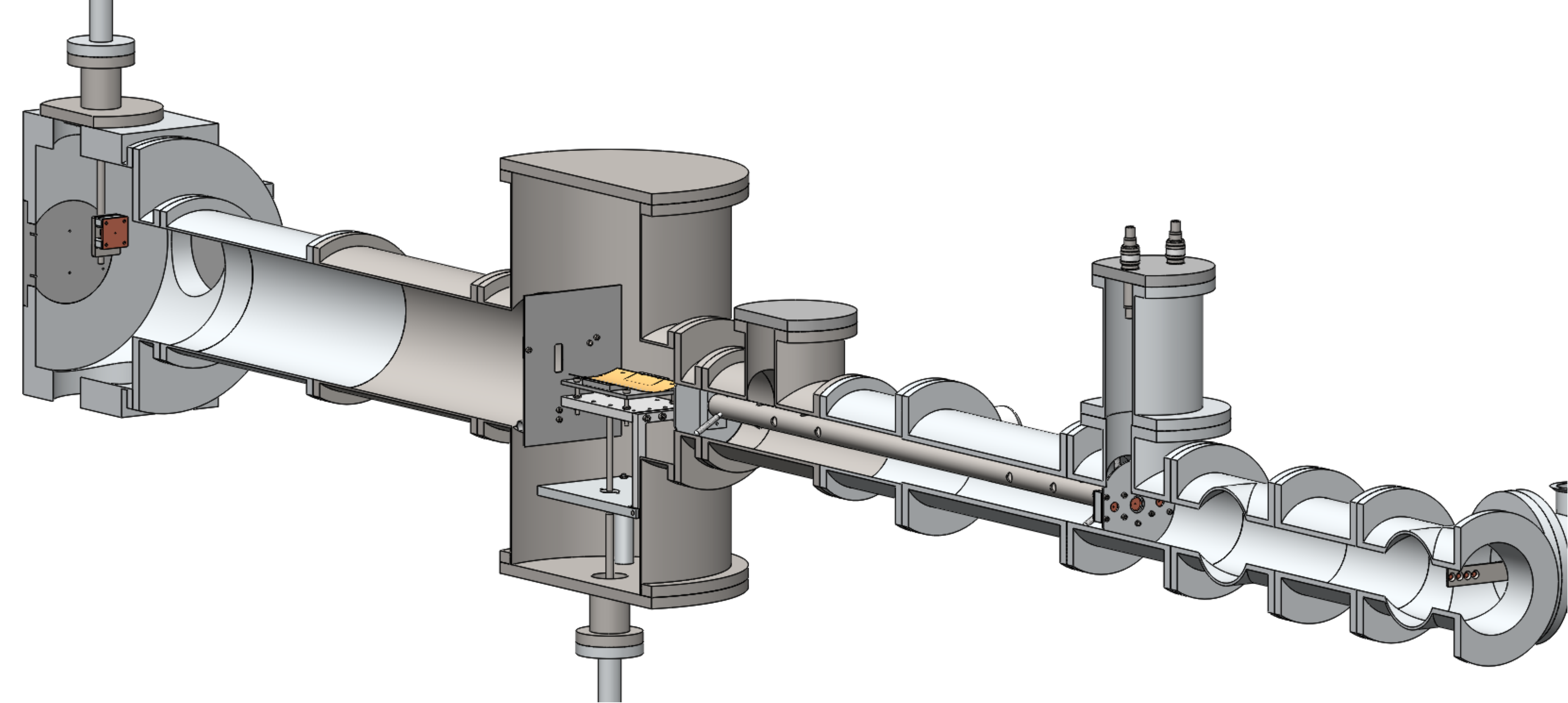}
	\caption{A CAD model of the scattering setup in cross-sectional view showing, from right to left, the degrader foils on a movable holder, the collimator with the monitor detectors around the entrance, the tiltable target table, and the shiftable detector for scattered protons at the end of the beam line.}
	\label{fig:cad}
\end{figure*}

A schematic of the setup is presented in Fig.~\ref{fig:scheme}, while the CAD model in cross-sectional view in Fig.~\ref{fig:cad} provides an insight into the implementation. Two angles are used in the following to describe the scattering geometry: the incidence angle $\Psi$, defined between mirror surface and incident beam, and the scattering angle $\Theta$, which is the angle of the detector position with respect to the incident beam.

The proton beam enters the setup through a pair of slits, separated about 50\,cm. Each of these slits consists of four adjustable copper parts that form a rectangular aperture. The individual parts are electrically isolated to allow the tapping of currents. The proton beam is well aligned with the setup when the currents on all opposite slit parts are equal.

A pinhole aperture reduces the incoming beam current and, therefore, the rates on the detectors. Depending on the actual measurement configuration, typical diameters are 0.1\textendash 1.0\,mm to prevent excessive pile-up while keeping measurement durations reasonable. As well as all other apertures in the setup, the pinhole is made of copper and has a chamfer on the downstream side to minimize scattering inside the aperture. Different geometries have been studied by means of Monte Carlo simulations with the \textit{Geant4} toolkit. The results have been incorporated in the actual design and can be found in \cite{Diebold_2014}.

The main experimental challenge is the determination of the incident particle flux during a scattering measurement. The beam intensity is subject to unavoidable fluctuations, which arise from variations of the terminal voltage and the yield of the ion source as well as, on longer timescales, from a drifting of the magnetic fields of the bending magnets. In order to normalize the measured scattering rates, monitor detectors are placed peripheral to the beam. A metal degrader foil of a few micrometers thickness is mounted downstream of the pinhole aperture to widen the beam spatially, providing a constant fraction of flux to these detectors. The ratio of target flux to the readings of the monitor detectors is determined in dedicated measurements. Since the monitors are mounted relatively close to the degrader and the flux of the widened beam decreases with distance, rather small 60\,\textmu m apertures in front of the detectors limit the rate to restrict the pile-up fraction to about 1\%. The overall accuracy of determining the target flux via off-axis monitor detectors is estimated from different measurements to be within $\pm 10\%$. An additional advantage of the degrader foils is the reduction of the beam energy, which makes energies below the lower limit of the accelerator accessible, whereas the drawback is a spectral broadening up to a few tens of keV.

\begin{figure}[tb]
	\centering
	\subfigure[Collimator exit]{
  	\includegraphics[width=.45\textwidth]{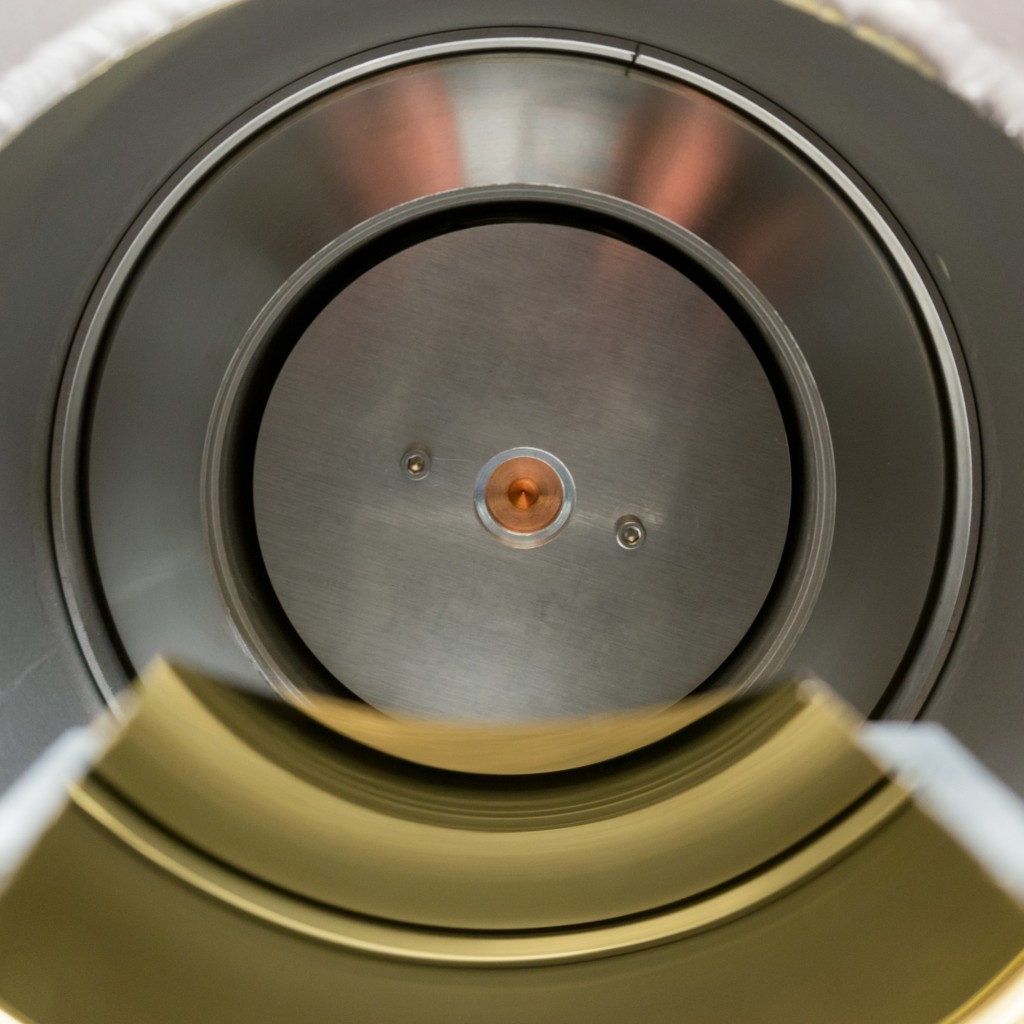}
		\label{fig:collimator}}
  \subfigure[Target mounting]{
  	\includegraphics[width=.45\textwidth]{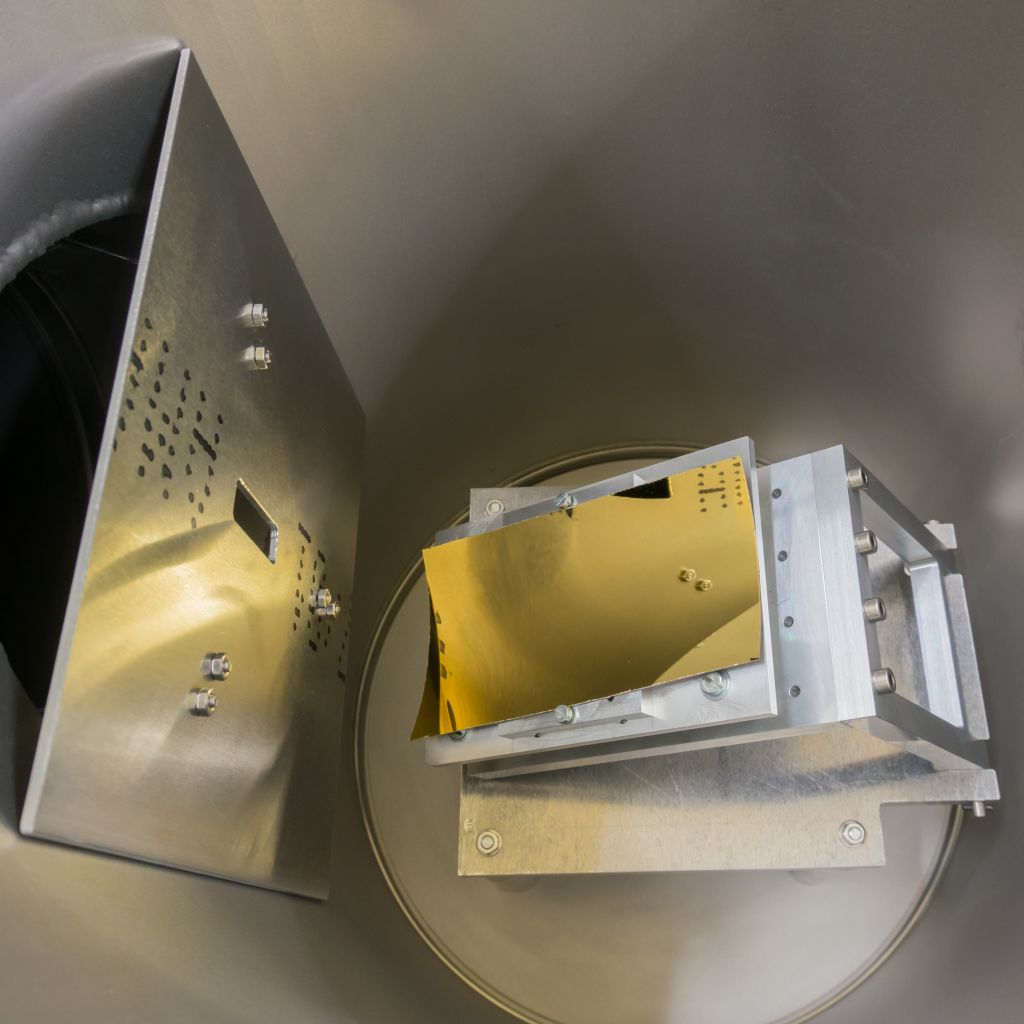}
		\label{fig:target}}
	\subfigure[Shiftable detector]{
  	\includegraphics[width=.45\textwidth]{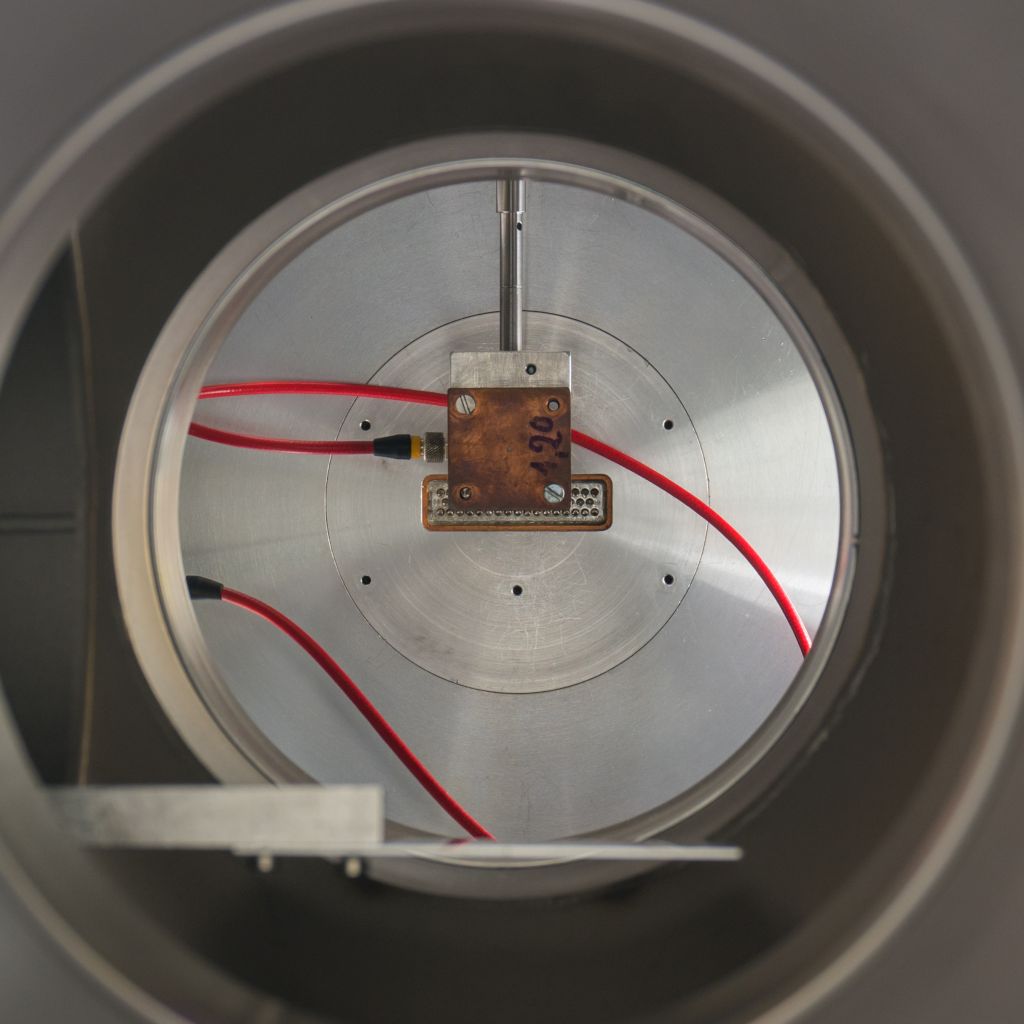}
		\label{fig:detector}}
	\subfigure[Detector at the collimator exit]{
  	\includegraphics[width=.45\textwidth]{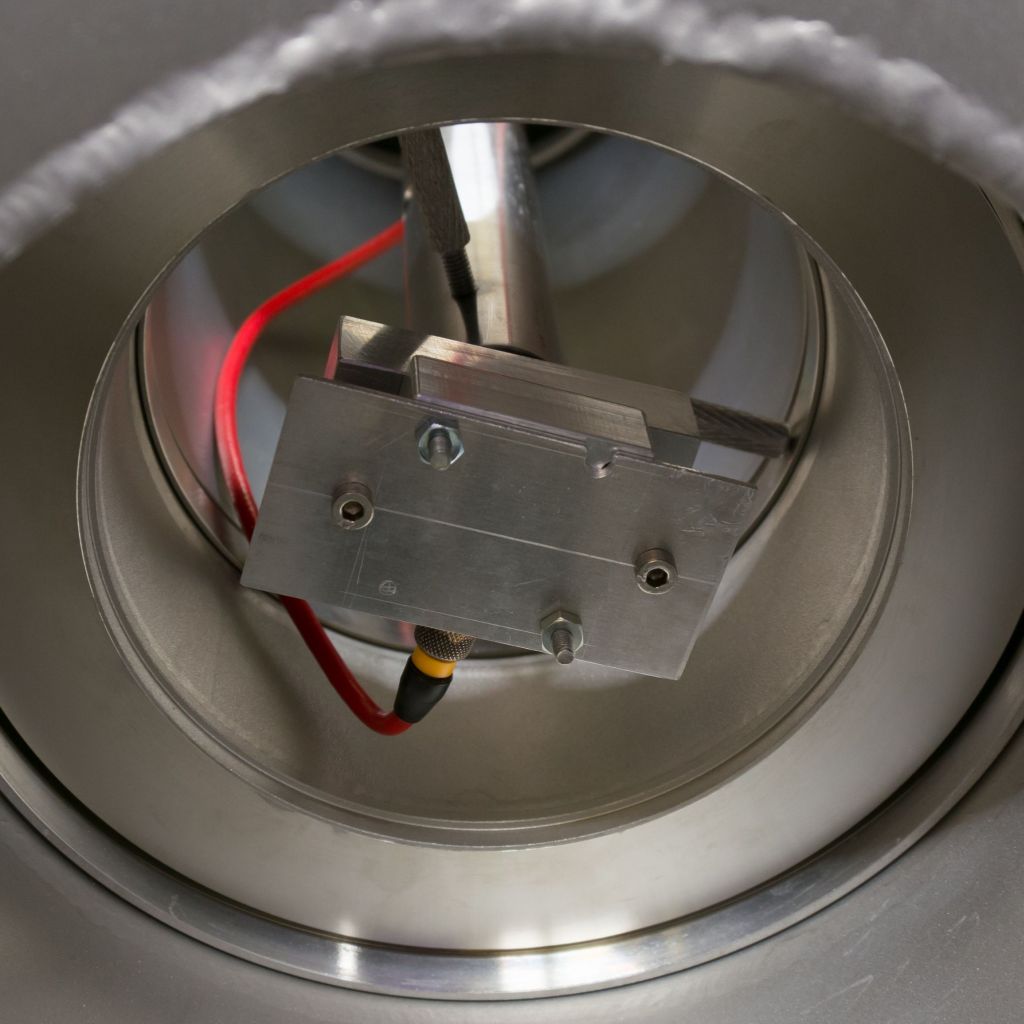}
		\label{fig:detColl}}
	\label{fig:close-ups}
	\caption{From top left to bottom right: Close-ups of the collimator exit that is surrounded with an aluminum plate to absorb bypassing protons; the tiltable target table with a part of an \textit{eROSITA} mirror shell and the aluminum sheet with a slit at the chamber exit, absorbing protons that have been scattered to larger angles; the movable detector for scattered protons at the end of the beam line; a detector mounted temporarily at the collimator exit to measure the transmitted proton flux.}
\end{figure}

In between the monitor detectors the entrance of a 78\,cm long collimator is located, which directs a part of the widened beam to the scattering target. During the measurements, aperture sizes of 1.0\,mm at the entrance and 0.3\,mm at the exit of the collimator have been used. This combination limits the maximum opening angle to 0.1\textdegree~(1.7\,mrad), providing a reasonable trade-off between transmitted flux and angular precision. The open space around the collimator tube is almost completely blocked at both ends with 2\,mm aluminum plates to absorb bypassing protons that have been scattered at the inner walls of the beam line (cf. Fig.~\ref{fig:collimator}). Instead of the second aluminum plate, a detector can be mounted temporarily on the exit of the collimator to determine the ratio of transmitted flux to the flux on the monitors in order to normalize the subsequent measurements (cf. Fig.~\ref{fig:detColl}).

The scattering target, i.e. a part of an X-ray mirror shell, is mounted on a tiltable table with a length of 120\,mm (cf. Fig.~\ref{fig:target}). Its height can be regulated via setscrews to adapt to different mirror geometries and thicknesses. Since the mirrors are curved, they are fixed with two screws on movable wedges that support them sideways. The screws have been tightened with minimal force to avoid any deformation of the mirror. The tilt angle, i.e. the incidence angle $\Psi$, is controlled via a linear manipulator from below. The precision of the manipulator of 0.01\,mm determines the theoretical precision of the tilt angle to 0.006\textdegree~(0.1\,mrad). The tilting axis is several centimeters below the beam, allowing to remove the target from the course of beam for a determination of the primary beam position on the detector plane. 

An aluminum sheet to reduce the flux of protons, which have been scattered at the inner walls of the beam line, is installed downstream of the target chamber (cf. Fig.~\ref{fig:target}). It leaves just a slit of 3\,cm height and 1\,cm width. A comparison of the proton background with and without this sheet shows a background reduction by about a factor of 50, which means a decisive increase of the signal to background ratio in a typical measurement scenario.

At a distance of 933\,mm to the center of the target table a detector is mounted on a linear manipulator of the same type as used for the mirror tilting (cf. Fig.~\ref{fig:detector}). The detector is vertically shiftable up to 75\,mm from the beam line center without shadowing effects, which corresponds to a maximal scattering angle $\Theta$ of about 4.5\textdegree~(80\,mrad). A 1.2\,mm aperture defines the solid angle $\Omega$ to about 1.3\,\textmu sr with respect to the mirror center. The diameter of this aperture has been determined by means of an x-y-table with micrometer position accuracy and an attached microscope with cross-hairs; the error of this method has been estimated to $\pm0.1\,\mathrm{mm}$.

\subsection{Scattering Targets}\label{ssec:targets}
The mirror shells of the \textit{eROSITA} telescope consist of nickel substrates with a 50\,nm thick gold layer coated on the inner surface. This composition is quite similar to the \textit{XMM-Newton} mirrors, except that their gold layer has a thickness of 150\textendash200\,nm. However, \textsc{TRIM} simulations conducted within this work showed that this difference does not affect the grazing angle proton scattering significantly. The substrate thicknesses of the \textit{eROSITA} mirrors range from 540\,\textmu m for the outer shells to 200\,\textmu m for the inner ones. For the scattering measurements, two samples have been cut from an \textit{eROSITA} spare mirror (shell 25, 270\,\textmu m substrate thickness) by means of a wire cutter. This method has been selected after the study of various alternatives because it leads to a minimum of debris particles on the samples and affects the gold coating only very close to the edges. The sizes of the samples differ slightly: the larger one has a trapezoid shape with a length of 12\,cm (cf. Fig.~\ref{fig:target}), the smaller one is almost rectangular with a length of about 10\,cm. Although all results presented within this publication have been obtained with the larger sample, data from the smaller one have been used to cross-check for consistency.

The mirror samples have been handled with care to avoid damage or contamination of the surface and minimal strain has been applied for the mounting to prevent deformations. However, the samples have not been kept under clean room conditions; dust particles on the surface have been removed regularly before closing the target chamber by spraying with nitrogen. A few small scratches have been present on the surface, but not in the central region where the proton beam has been scattered.


\subsection{Proton Detectors}\label{ssec:dettectors}
Silicon surface barrier (SSB) detectors with 8\,mm sensitive diameter are used for the detection of the scattered protons and as monitor detectors. These SSBs feature a detection efficiency of almost 100\% and an energy resolution around 10\textendash20\,keV. The maximal count rate is limited by the shaping time of the spectroscopic amplifiers; typically rates of up to $2 \cdot 10^5\,\mathrm{s^{-1}}$ can be achieved without sacrificing the spectral resolution.

The low energy threshold, above which proton events can be unambiguously identified, is defined by the thermal noise level and the level of electromagnetic noise that is picked up by the detectors and the readout electronics. Although noise levels well below 80\,keV could be reached in the scattering setup, the trigger threshold has been set to about 100\,keV to account for an increase at high rates.

The signals of all three detectors are digitized by a fourfold peak-sensing ADC (analog-to-digital converter). In order to correct for dead time, a pulser signal is split and input on the fourth ADC channel as well as on a scaler. The fraction of pulses registered by the ADC compared to the scaler value gives the conversion efficiency for all ADC channels.

An energy calibration of the SSBs has been performed in the scattering chamber of beam line 2 by measuring backscattered protons from different known targets. Before the calibration, the incident beam energy has been determined via the 992\,keV resonance of the \ce{^{27}Al}(p,$\gamma$)\ce{^{28}Si} reaction. The complete calibration procedure is described in detail in \cite{Diebold_2013}; the accuracy is estimated to be within $\pm 10\,\mathrm{keV}$.


\subsection{Alignment and angular calibration}\label{ssec:calibration}
The alignment of all apertures in the course of beam, i.e. the slits, the pinhole aperture, and entrance and exit of the collimator, is performed with a theodolite that is placed behind the end of the beam line. In order to allow the alignment procedure, the detector chamber has a flange on the back side and all apertures can be removed.

As described in Section \ref{ssec:blsetup}, the incidence angle $\Psi$ and the scattering angle $\Theta$ are adjusted via linear manipulators, necessitating a conversion between the manipulator readings and the angles. The scattering angle can be calculated from the geometry of the setup, if the manipulator position corresponding to $\Theta = 0$\textdegree~is known, whereas the incidence angle needs to be calibrated. Furthermore, the entire incident flux encounters the mirror only within a certain range of incidence angles, depending on the height of the target table. The limits need to be determined, since a measurement of the scattering efficiency is only meaningful within this range.

Since the X-ray mirror targets reflect visible light and the SSB detector is sensitive to it as well, a 445\,nm solid-state laser is used for the calibration. The device is sufficiently compact to be inserted into the beam line after the pinhole. The position and orientation of the laser is adjusted via setscrews until the transmitted intensity through the collimator is maximized. After the $\Theta = 0$\textdegree~position of the detector is determined, the mirror is shifted in the course of beam. The position at which the primary beam is completely blocked corresponds to the minimal incidence angle. Successively, the tilt of the mirror is increased  and the position of the reflected beam measured with the detector. The corresponding angles are calculated via the condition $\Psi = \Theta / 2$ for specular reflection. During the calibration, the proper positioning of the mirror is checked as well. Therefore, the measured laser intensity is observed while slightly shifting the detector perpendicular to the manipulator direction. If the mirror surface, where the beam is reflected, is not aligned perpendicular to the manipulator direction, the reflected laser spot drifts away from the detector with increasing incidence angle. In this case, the mirror position is changed accordingly and the calibration procedure is repeated. This procedure has been iterated until any observable drift has vanished.

\begin{figure}
  \includegraphics[width=.75\textwidth]{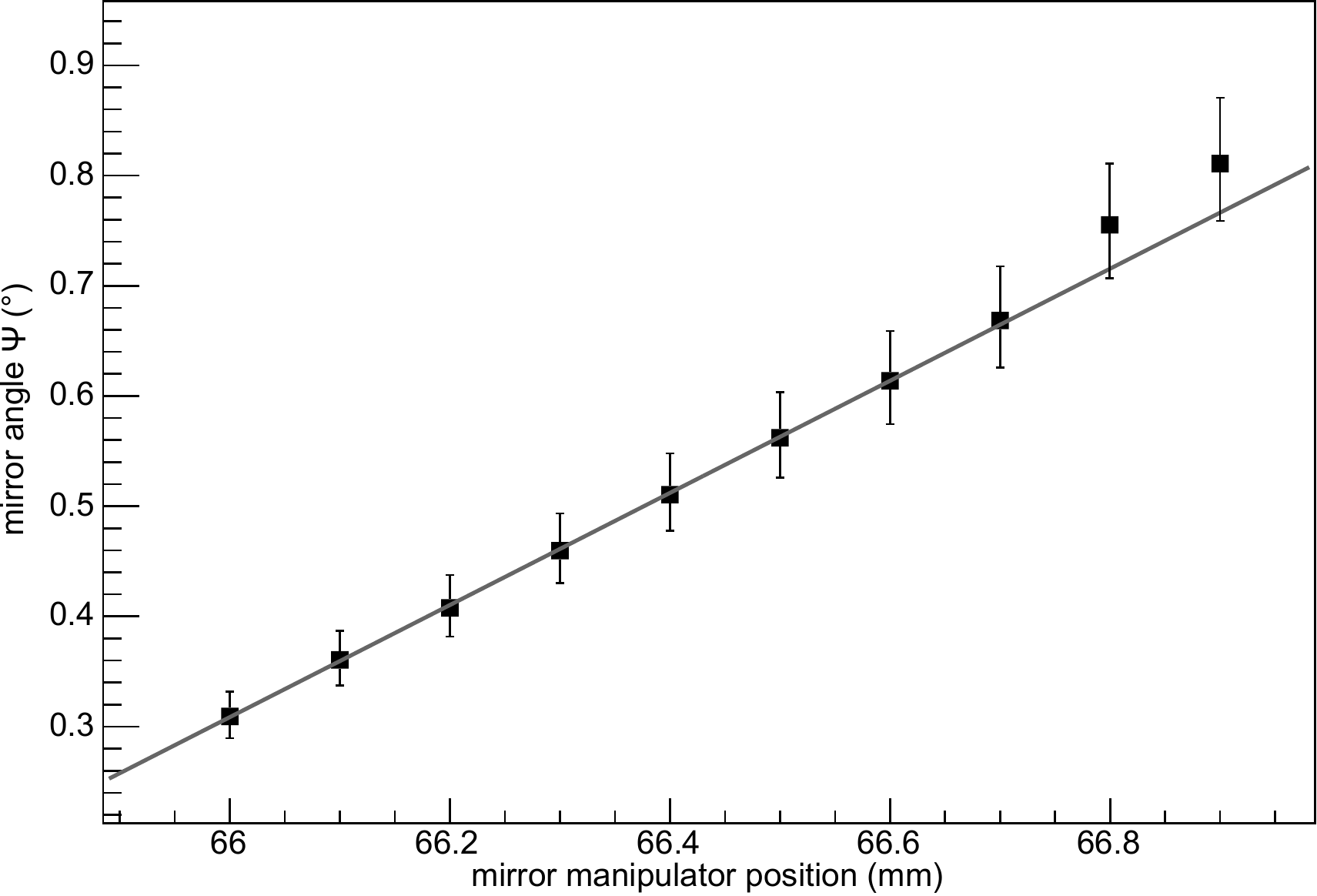}
	\caption{Plot of the laser calibration data for one of the \textit{eROSITA} mirror samples. The grey line is a linear fit to the data from 66.0 to 66.7\,mm. The points of larger positions are shifted because the incident beam hits the front of the mirror sample. This marks the maximal incidence angle for a meaningful measurement of the scattering efficiency.}
	\label{fig:mircal}
\end{figure}

As an example, laser calibration data for one of the \textit{eROSITA} mirror samples is shown in Fig.~\ref{fig:mircal}. The first data point at a mirror manipulator position of 66.0\,mm marks the point at which the mirror blocks completely the primary laser beam. Thereafter, the correlation with the detector position is linear up to a manipulator reading of 66.7\,cm, where a shift of the reflected beam is visible. This shift appears because the beam starts to hit the front of the mirror sample and marks the maximal mirror angle for this setting.

\section{Measurement Parameters and Procedures}\label{sec:meas}

\subsection{Proton Energies and Spectra}
\begin{figure}
  \includegraphics[width=.8\textwidth]{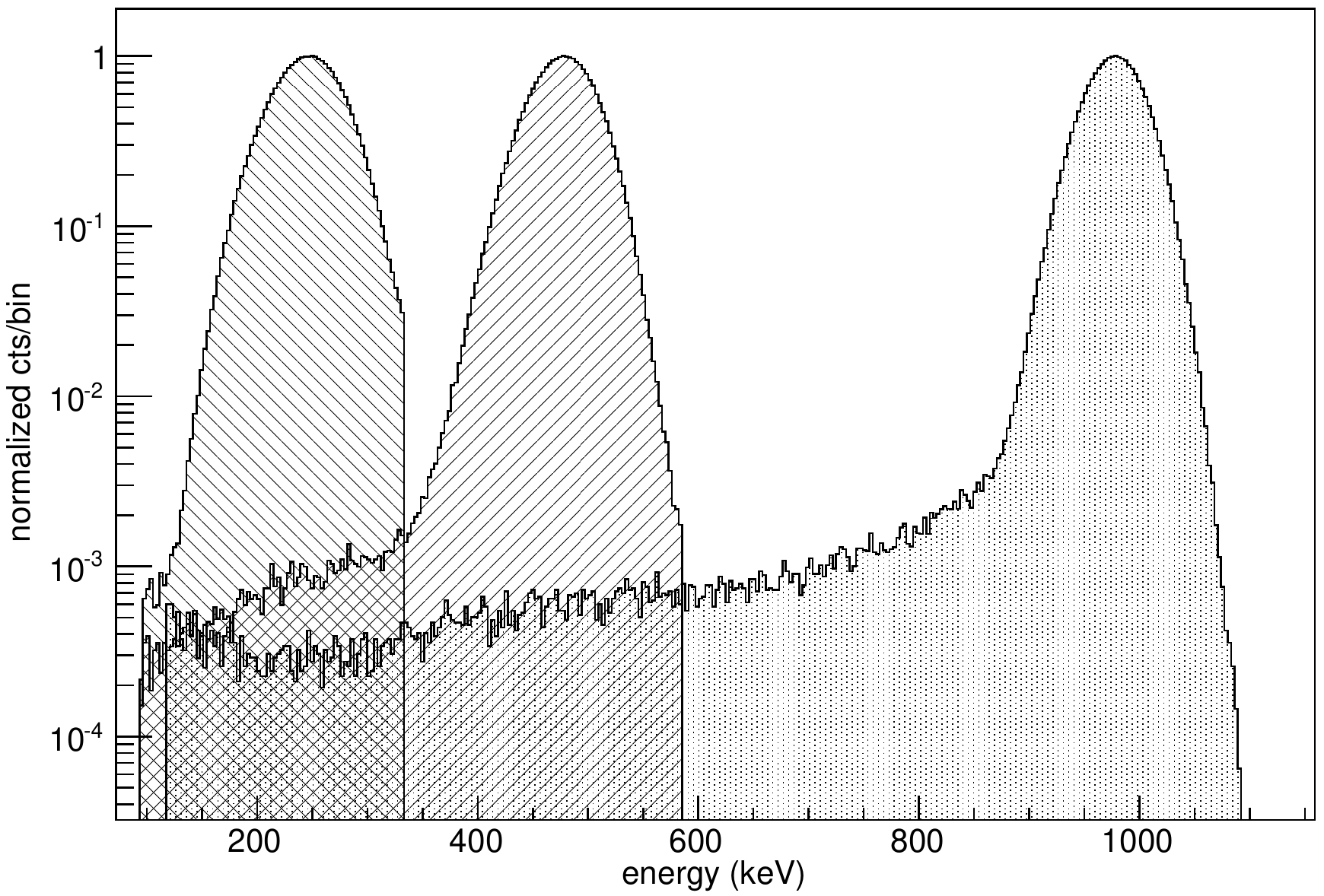}
	\caption{Overlay of the incident proton spectra, recorded with the detector at the end of the beam line. The spectra are centered around 245\,keV, 477\,keV, and 977\,keV. The low energy tails are mostly due to slit scattering. Pile-up has been removed by applying an energy threshold cut.}
	\label{fig:spectra}
\end{figure}

Three proton energies have been selected for the scattering measurements: 250\,keV, 500\,keV, and 1\,MeV.  This range covers, as far as experimental constraints allow, the energies for which an enhanced funneling efficiency has been measured in orbit. Energies lower than 250\,keV are not reasonable in the current setup considering the 100\,keV detection threshold of the SSBs and a possible energy loss of several tens of keV. Combinations of initial beam energy, degrader material, and foil thickness to obtain the mentioned energies are derived from simulations by means of the \textsc{TRIM} code. Table~\ref{tab:energies} lists the combinations finally used, together with mean energy and FWHM that have been derived by fitting a Gaussian to the main part of the measured spectra (cf. Fig.~\ref{fig:spectra}). The Gaussian shape arises from straggling in the degrader foil and, to a smaller extend, from the finite energy resolution of the SSB detector. The low energy tails visible in the 500\,keV and 1\,MeV spectra are mostly due to slit scattering and contain at maximum 1\% of the total flux. Dedicated low flux measurements with a smaller pinhole aperture show that up to about 5\% of the counts in a spectrum of the incident beam are pile-up; a high energy tail above the Gaussian peak is not present in the data. Since pile-up is unavoidable when detecting the incident beam directly under typical measurement conditions, an energy threshold cut has been applied and all recorded events above the threshold have been counted twice. The same procedure has been applied in the analysis of the scattering data.

Although the mean energy incident on the target $E_\mathrm{inc}$ shifts within $\pm20\,\mathrm{keV}$ between individual measurements, the nominal energies are given in the following for the sake of convenience. The deviation does not affect the results of the efficiency measurements; in the analysis of the energy loss it has been accounted for.

\begin{table}
\caption{List of the beam energy and degrader combinations used for the scattering measurements. The spectral parameters of the exiting beam have been determined by fitting Gaussians to the main part of the measured spectra (cf. Fig.~\ref{fig:spectra}).}
\label{tab:energies}
\begin{tabular}{cccc}
\hline\noalign{\smallskip}
Beam energy & Degrader material & Mean energy & FWHM \\
(keV) & and thickness & (keV) & (keV) \\
\noalign{\smallskip}\hline\noalign{\smallskip}
1030 & 6\,\textmu m Cu & 245 & 76 \\
1180 & 6\,\textmu m Cu & 477 & 70 \\
1550 & 6\,\textmu m Cu & 977 & 64 \\
\noalign{\smallskip}\hline
\end{tabular}
\end{table}

\subsection{Scattered Proton Spectra}
\begin{figure}
  \includegraphics[width=.7\textwidth]{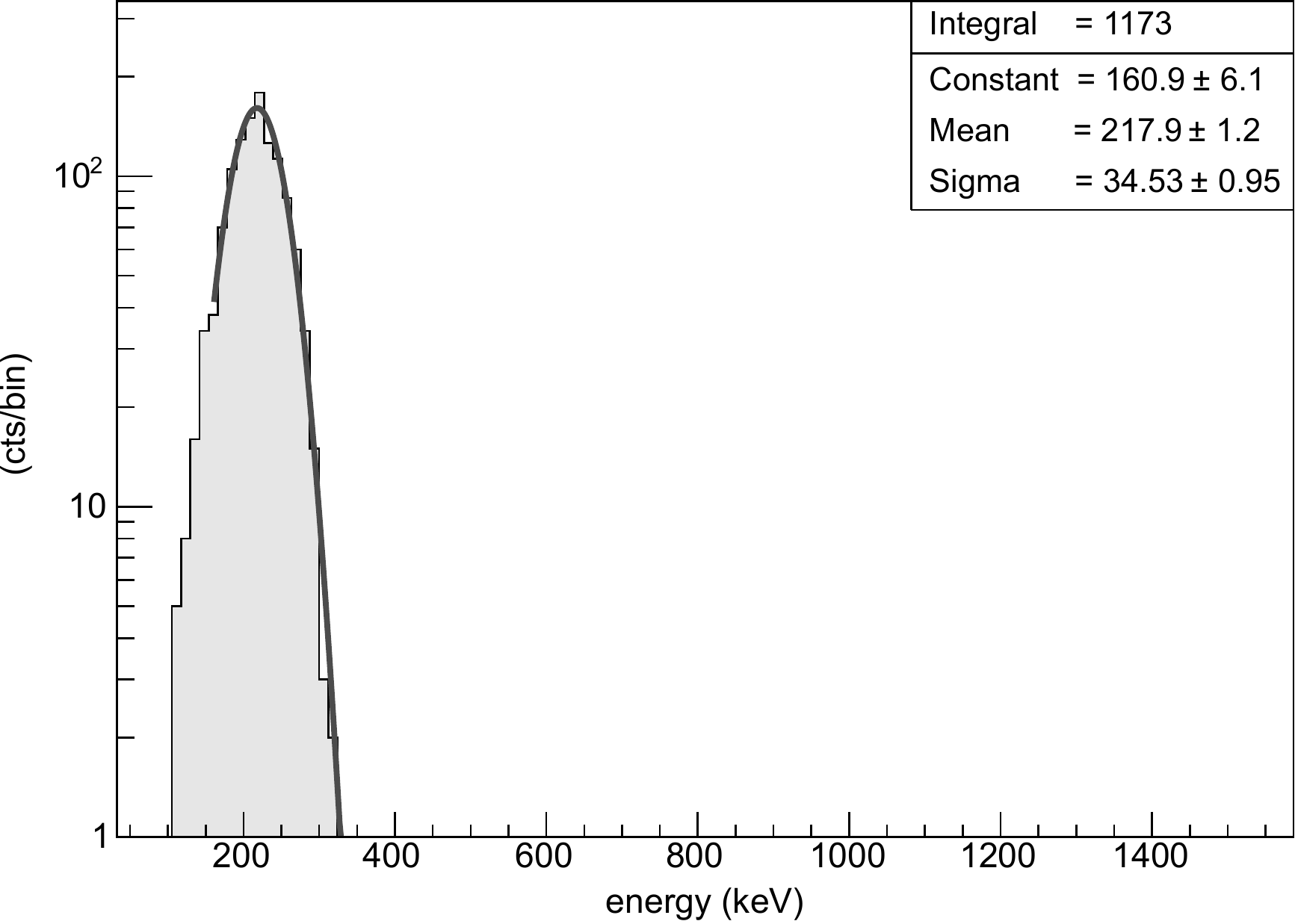}
	\caption{Spectrum of scattered protons for $E_\mathrm{inc}=246.7\,\mathrm{keV}$, incidence angle $\Psi=0.67^\circ \pm 0.11^\circ$, and scattering angle $\Theta=1.67^\circ \pm 0.11^\circ$. The mean energy loss has been determined by fitting a Gaussian to the main peak (grey line) and comparing the position with a Gaussian fit to the incident spectrum.}
	\label{fig:spec250keV}
\end{figure}

\begin{figure}
  \includegraphics[width=.7\textwidth]{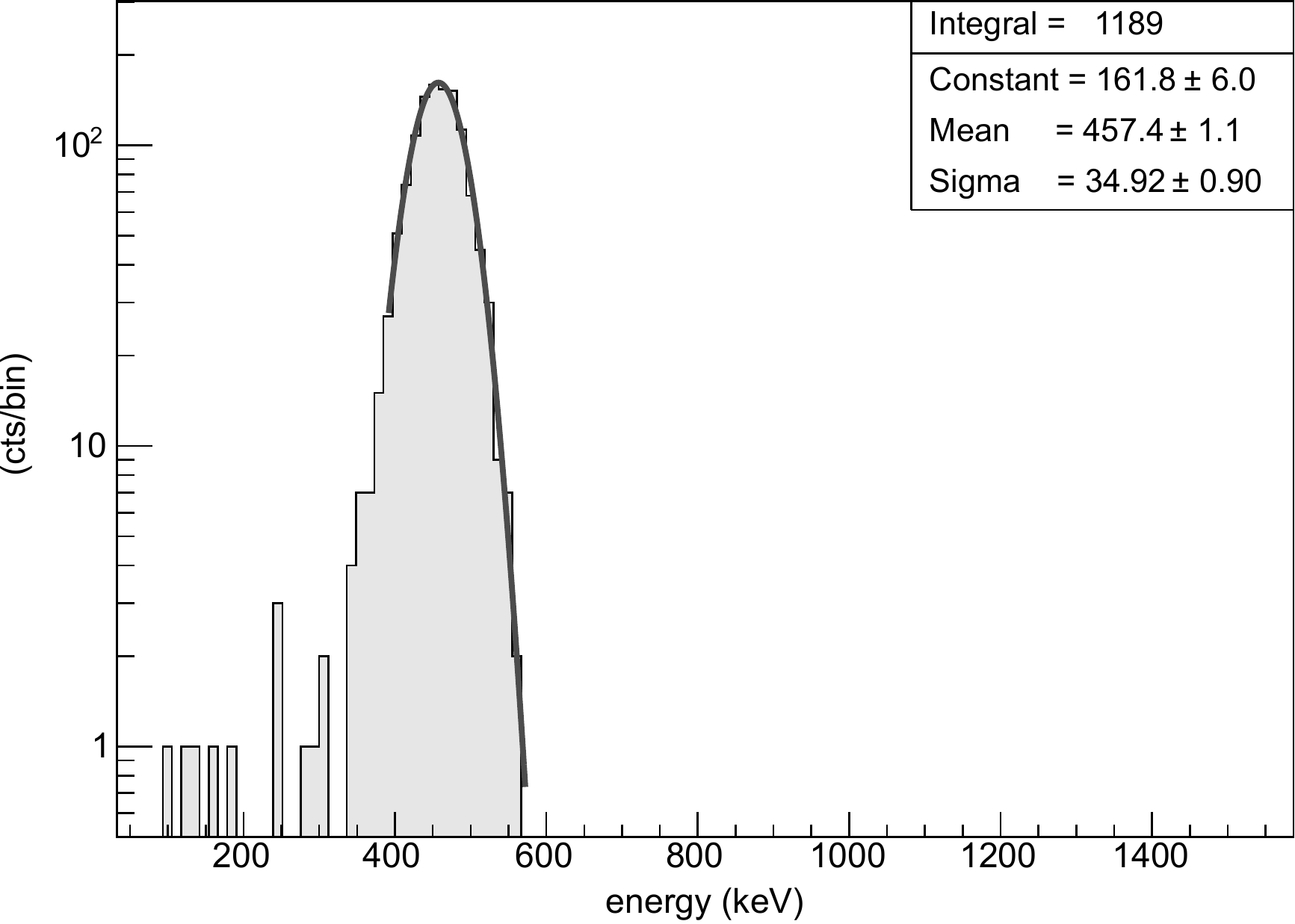}
	\caption{Same as Fig.~\ref{fig:spec250keV}, but for $E_\mathrm{inc}=492.9\,\mathrm{keV}$, $\Psi=0.64^\circ \pm 0.12^\circ$, and $\Theta=1.64^\circ \pm 0.11^\circ$.}
	\label{fig:spec500keV}
\end{figure}

\begin{figure}
  \includegraphics[width=.7\textwidth]{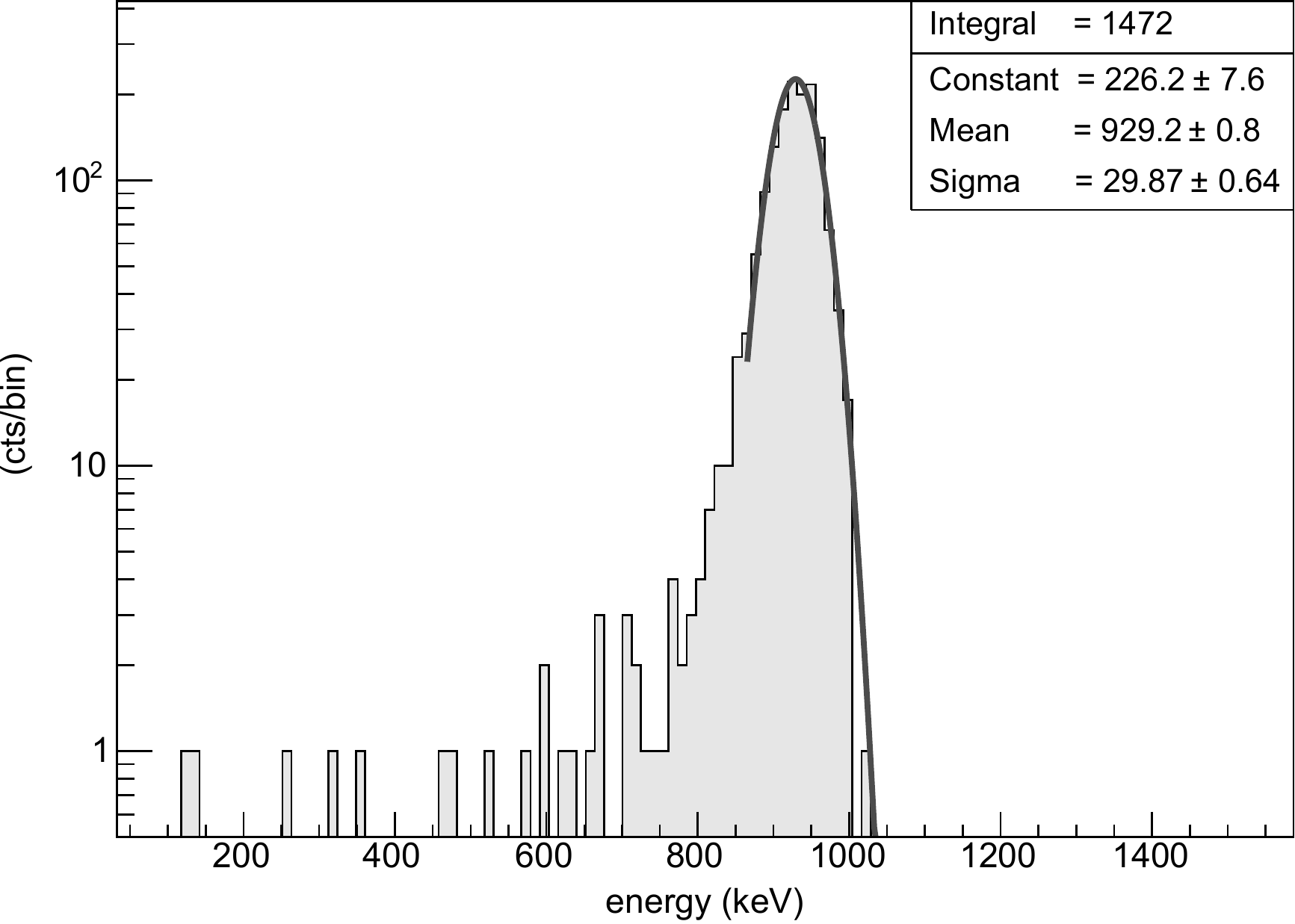}
	\caption{Same as Fig.~\ref{fig:spec250keV}, but for $E_\mathrm{inc}=968.4\,\mathrm{keV}$, $\Psi=0.61^\circ \pm 0.15^\circ$, and $\Theta=1.61^\circ \pm 0.10^\circ$.}
	\label{fig:spec1MeV}
\end{figure}

Examples of scattered proton spectra at mean incidence and scattering angles are given in Figs.~\ref{fig:spec250keV}\textendash\ref{fig:spec1MeV} for the three incident energies. Almost all acquired spectra contain at least 1000 counts. While for the calculation of the scattering efficiency only the integrals have been considered, Gaussians have been fitted to determine the position of the main peak for a calculation of the mean energy loss $\Delta E$ by a comparison with the incident spectrum. In order to account for a possible drift of the primary beam energy, the incident spectra have been measured frequently during a set of measurements.

\subsection{Beam Direction and Incident Flux}
\begin{figure}
  \includegraphics[width=.7\textwidth]{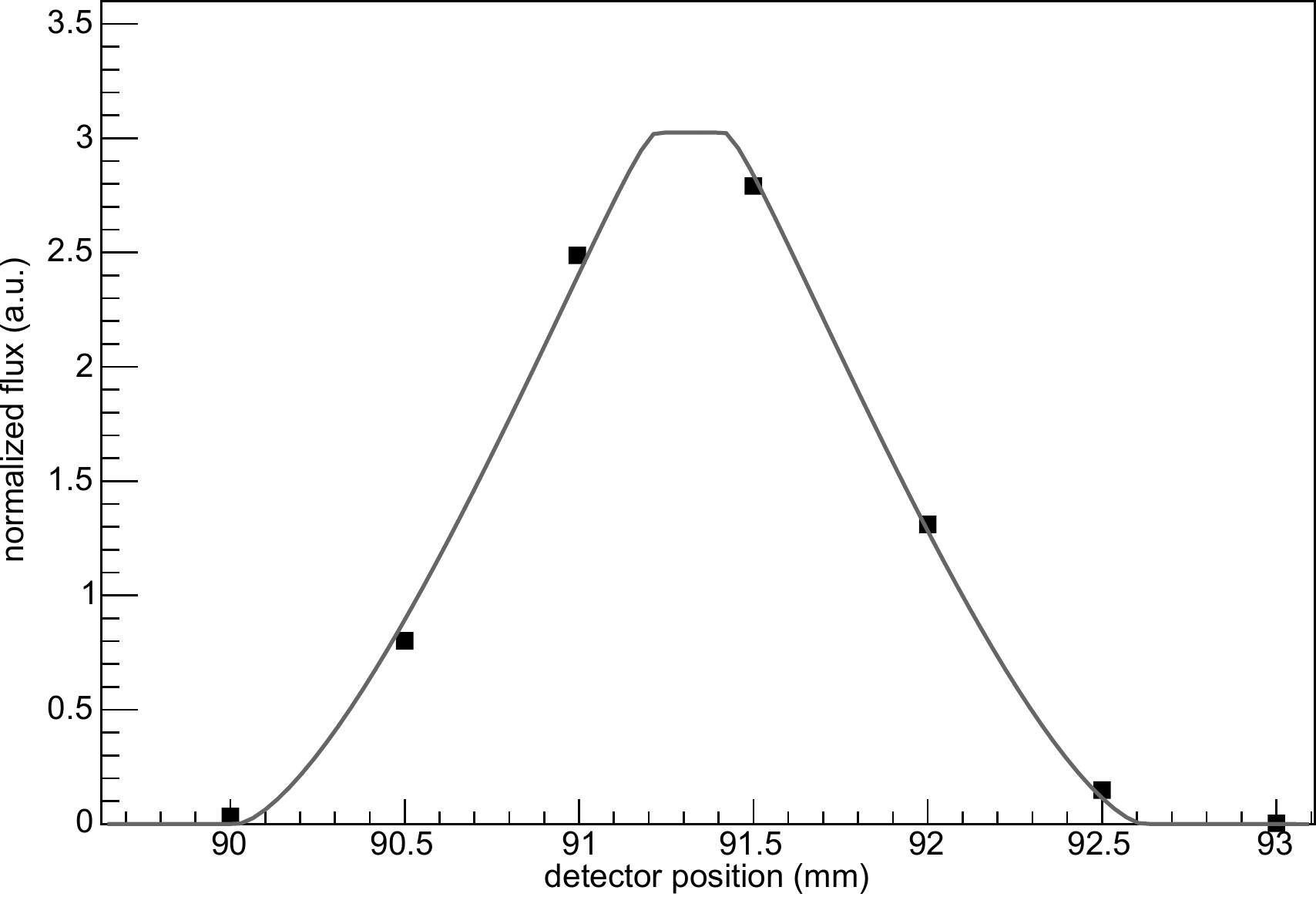}
	\caption{Plot of the proton flux around the primary beam position ($\Theta = 0$\textdegree). The mirror has been tilted out of the course of beam for this measurement. A function that describes the area of two partially overlapping circles has been fitted to the data (grey line).}
	\label{fig:pBeamPos}
\end{figure}

The position of the incident proton beam at the detector plane varies on the order 0.5\textendash1\,mm between sets of measurements, when the beam has been set up newly, and deviates up to 1.5\,mm from the position measured during the laser calibration. Therefore, the mirror has been tilted out of the course of beam at the beginning of each set and the flux at different position close to the primary beam ($\Theta = 0$\textdegree) has been measured. The primary beam position has been derived from this data via fitting with a function that describes the overlapping area of two circles with different radii in dependence of the distance of the circle centers. Data of such a measurement and its good agreement with the theoretical description is shown in Fig.~\ref{fig:pBeamPos}. The actual primary beam position has been considered in the final calculation of both angles, $\Psi$ and $\Theta$.

The variation of the direction of the incident proton beam impedes the determination of the incident flux from the measurement at the primary beam position, since the deviation is of the order of the diameter of the detector aperture and the detector can only be moved vertically. As an alternative, a detector has temporarily been mounted at the collimator exit to determine the transmitted proton flux (cf. Fig.~\ref{fig:detColl}). A normalization parameter has been derived from this measurement, which associates the measured flux on the monitor detectors to the incident flux on the target. Since the widening of the beam by the degrader foils is energy dependent, this measurement has been repeated for all energies used. All measured fluxes have been normalized to the mean flux of the two monitor detectors. The uncertainty for the incident flux on the target by applying this method has been estimated to $\pm 10\%$.

Furthermore, the proton background at each of the measured scattering angles has been determined while the mirror has been tilted out of the course of beam. Background arises because protons scatter at the exit aperture of the collimator and hit~\textendash~either directly or via a second small angle scatter on the inner beam line wall~\textendash~the detector. The background contribution is on the order of a few percent and has been subtracted during the analysis of the scattering efficiency.


\subsection{Angular Coverage}
The intention of the measurements is to cover the angles relevant for the assessment of the proton funneling to the focal plane of Wolter type-I mirror assemblies, which in principle comprises incidence angles in the range 0\textdegree\textendash90\textdegree. Nevertheless, since the efficiency drops significantly towards larger incidence angles, the measurements have been focused on grazing incidence angles. The minimal feasible mirror angle in the experiment is about 0.3\textdegree, therefore incidence angles between this lower limit and 1.2\textdegree~have been investigated in combination with scattering angles in the range 0.5\textdegree~to 4.1\textdegree.


\section{Results and Discussion}\label{sec:results}

\subsection{Scattering Efficiency}\label{ssec:eff}
The differential scattering efficiency $\eta$ is calculated according to Eq.~\ref{eq:eff} by dividing the detected number of protons $N_\mathrm{det}$ for a specific incidence angle $\Psi$ and scattering angle $\Theta$ by the number of incident protons $N_\mathrm{inc}$ on the target mirror. The result has been normalized to the solid angle $\Omega$ of the detector, considering the dependence on $\Theta$.

\begin{equation}
	\eta(\Psi,\Theta) = \frac{\Phi_\mathrm{det}(\Psi,\Theta)}{\Phi_\mathrm{inc}}\frac{1}{\mathrm d \Omega} = \frac{N_\mathrm{det}(\Psi,\Theta)}{N_\mathrm{inc}} \frac{1}{\Omega(\Theta)}
\label{eq:eff}
\end{equation}

\begin{figure}
  \includegraphics[width=.8\textwidth]{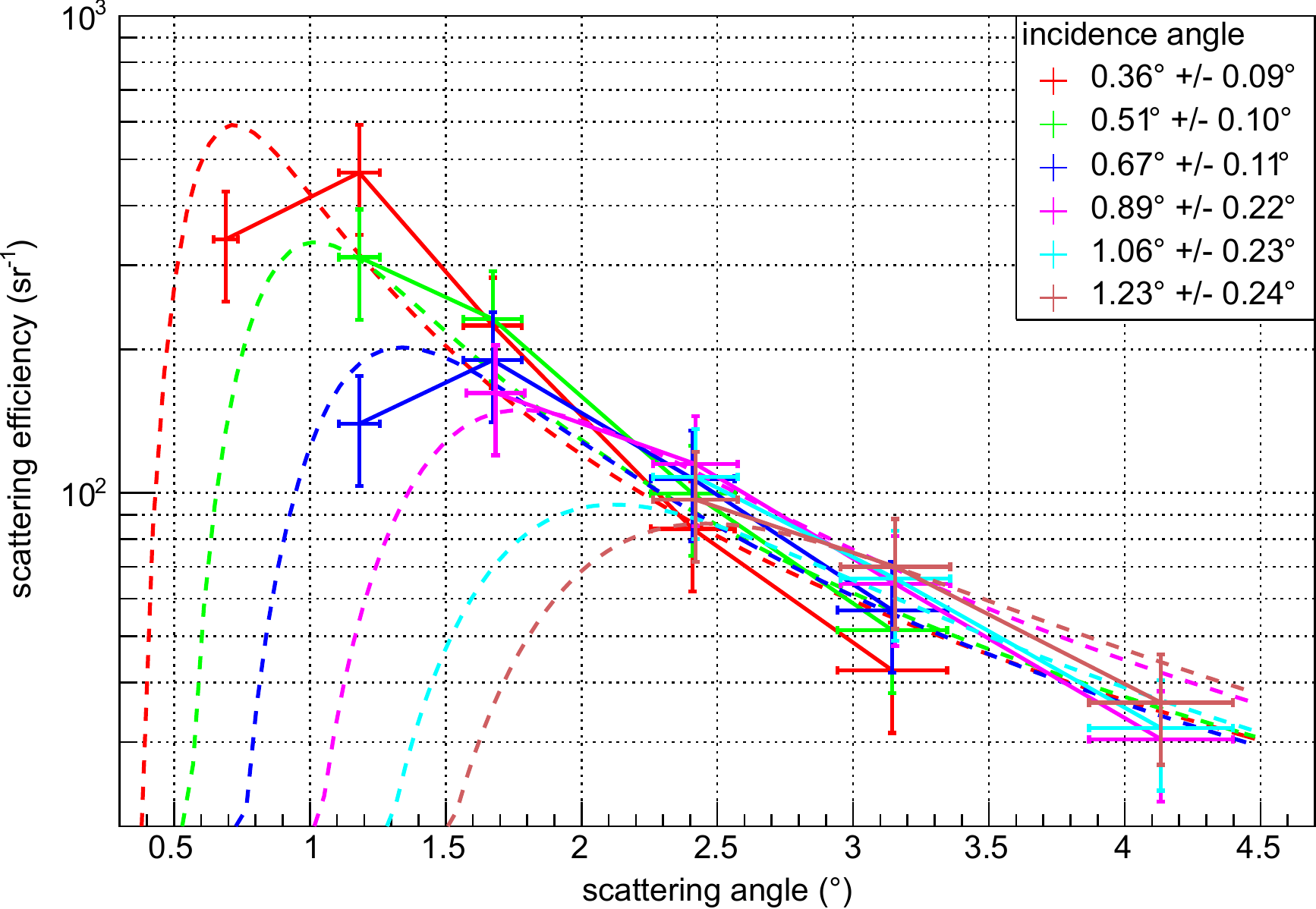}
	\caption{Scattering efficiency results for $E_\mathrm{inc} = 250\,\mathrm{keV}$. The values are normalized to the solid angle of the detector. The given errors account for counting statistics as well as uncertainties of the solid angle and the incidence and scattering angles. The individual data points for each incidence angle are connected by straight lines. The dashed lines are calculated by means of the Firsov formula (cf. Section~\ref{ssec:comparison}).}
	\label{fig:plot250keVeff}
\end{figure}

\begin{figure}
  \includegraphics[width=.8\textwidth]{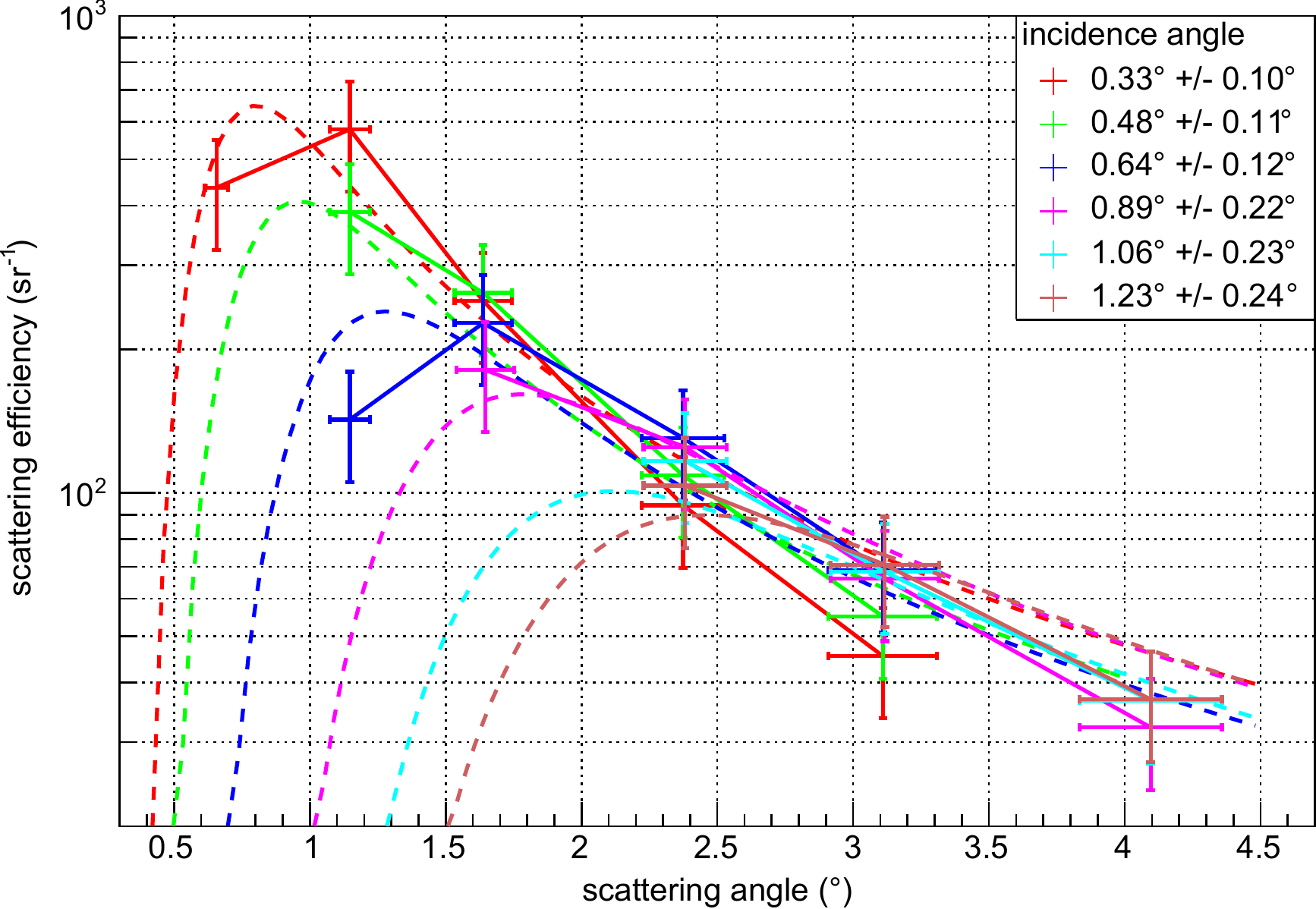}
	\caption{Same as Fig.~\ref{fig:plot250keVeff}, but for $E_\mathrm{inc} = 500\,\mathrm{keV}$.}
	\label{fig:plot500keVeff}
\end{figure}

\begin{figure}
  \includegraphics[width=.8\textwidth]{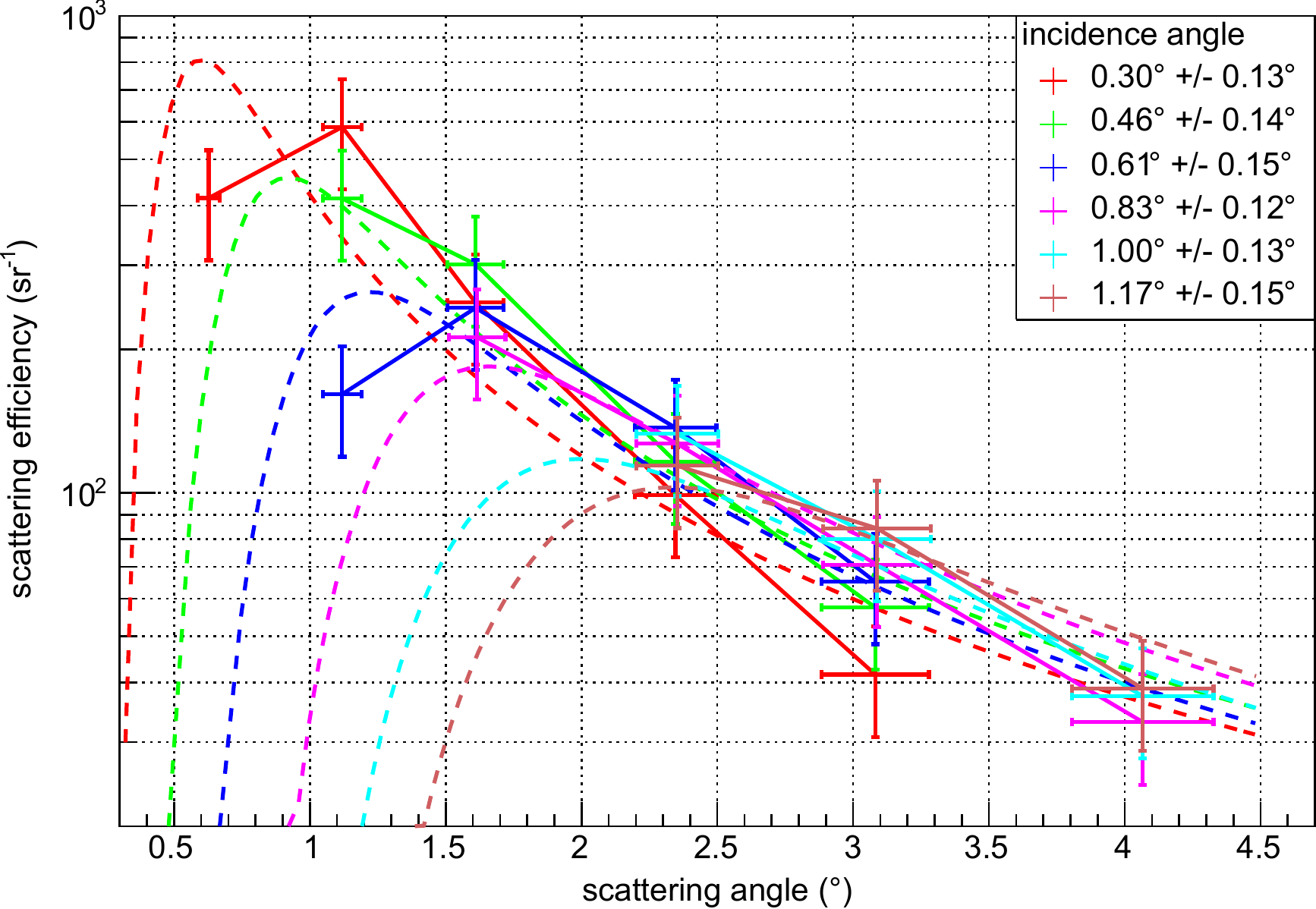}
	\caption{Same as Fig.~\ref{fig:plot250keVeff}, but for $E_\mathrm{inc} = 1\,\mathrm{MeV}$.}
	\label{fig:plot1MeVeff}
\end{figure}

The experimentally determined scattering efficiencies for the three incident energies $E_\mathrm{inc}$ are presented in Figs.~\ref{fig:plot250keVeff}\textendash\ref{fig:plot1MeVeff}. The data have been acquired with two different settings of the target mirror, one optimized for $\Psi<0.7$\textdegree, the other for $\Psi>0.7$\textdegree. As the mirror bulk is shielding the detector for $\Psi > \Theta$, these data points are compatible with zero and, therefore, have not been plotted.

The errors for the scattering efficiency shown in the plots comprise counting statistics (effect on efficiency $<5\%$), the estimated uncertainties of the effective detector area (19\% in excess and 15\% in deficit) as well as of the incident flux determination by using the peripheral monitor detectors ($\sim \pm 10\%$), the calibration and mechanical accuracy of incidence and scattering angles, and the uncertainty of the distance between the detector and the actual scattering position on the mirror (12\% in excess and 15\% in deficit). The latter dominates the angular error and causes the increase towards larger scattering angles because~\textendash~besides the inevitable extension of the beam spot to several centimeters~\textendash~the position of the spot on the 12\,cm long target cannot be localized.

In order to check for systematic effects introduced by the mirror mounting and the laser calibration procedure, the measurements for $\Theta<0.7$\textdegree~have been repeated twice: once with the same mirror mounted reversely and once with the second mirror sample (cf. Section~\ref{ssec:targets}). The data of all three measurements are consistent within the given errors.

The results show slightly higher efficiencies for larger incident energies. Table~\ref{tab:effcomp} lists the maximum measured efficiency $\hat\eta$ and the corresponding scattering angle for each incident energy and incidence angle. The maximum is in all measurements found around $\Theta = 2\Psi$, the condition for specular reflection.

\begin{table}
\caption{For each incidence angle $\Psi$ and incident energy $E_\mathrm{inc}$, the maximum efficiency $\hat\eta$ and the corresponding scattering angle $\Theta$ is listed. Since incidence and scattering angle deviate slightly between individual measurements, the mean values are given.}
\label{tab:effcomp}
\begin{tabular}{ccccccc}
\hline\noalign{\smallskip}
$\Psi_\mathrm{mean}$ & $\Theta_\mathrm{mean}$ & $\hat\eta_\mathrm{250\,keV}$ & $\hat\eta_\mathrm{500\,keV}$ & $\hat\eta_\mathrm{1\,MeV}$ \\
(\textdegree) & (\textdegree) & ($\mathrm{sr}^{-1}$) & ($\mathrm{sr}^{-1}$) & ($\mathrm{sr}^{-1}$) \\
\noalign{\smallskip}\hline\noalign{\smallskip}
0.33 & 1.15 & $469 \pm 122$ & $578 \pm 150$ & $585 \pm 152$ \\
0.48 & 1.15 & $312 \pm  81$ & $388 \pm 101$ & $414 \pm 107$ \\
0.64 & 1.64 & $190 \pm  49$ & $227 \pm  59$ & $244 \pm  63$ \\
0.86 & 1.65 & $162 \pm  42$ & $181 \pm  47$ & $212 \pm  55$ \\
1.03 & 2.38 & $108 \pm  28$ & $117 \pm  30$ & $133 \pm  35$ \\
1.20 & 2.38 & $ 97 \pm  25$ & $104 \pm  27$ & $114 \pm  30$ \\
\noalign{\smallskip}\hline
\end{tabular}
\end{table}

These results deviate from the findings in \cite{XMM-RGS} for \textit{XMM-Newton} mirrors, only the present data for 250\,keV are in agreement with the reported efficiencies for $E_\mathrm{inc}=300\,\mathrm{keV}$. For 500\,keV the efficiencies of the present study are almost a factor of two higher; a comparison of the present 1\,MeV measurement to the reported 1.3\,MeV \textit{XMM-Newton} data yields an even larger discrepancy around a factor of three and four. However, it should be stated that only six data points could be compared directly. The target mirrors and the experimental setups are comparable, except for the method of determining the incident flux, which might explain at least partially the systematically lower efficiencies of the former results towards higher energies: the current on the target mirror in the \textit{XMM-Newton} setup has been integrated with an electrometer; from the collected charge the number of incident protons has been calculated. Although the emission of secondary electrons from the target has been reduced with a repulsion grid, the yield of these delta electrons is increasing with rising beam energy, possibly leading to an overestimation of the collected charge at higher energies and, therefore, to apparently lower efficiencies.

\subsection{Energy Loss}
\begin{figure}
  \includegraphics[width=.8\textwidth]{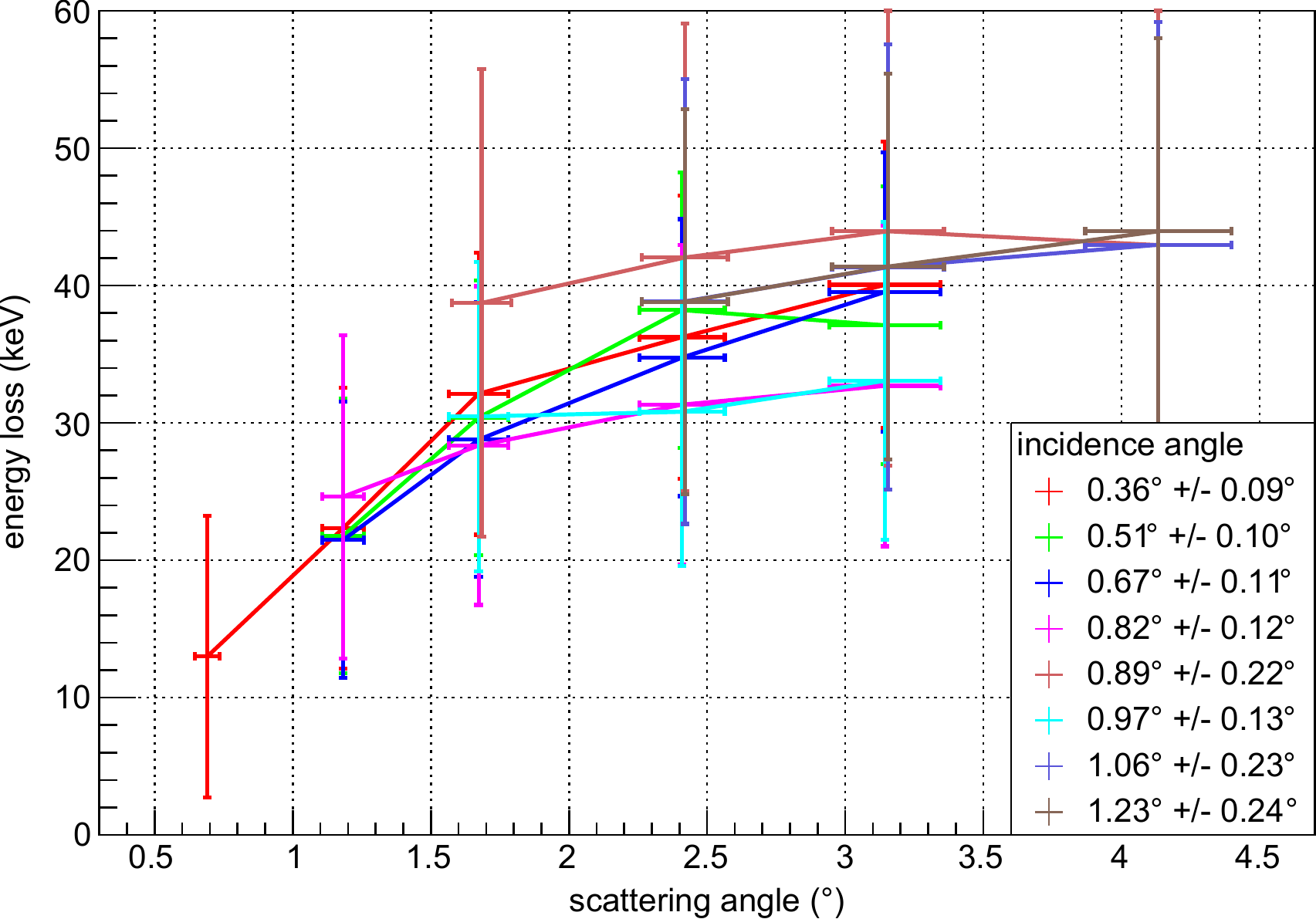}
	\caption{Most probable energy loss for $E_\mathrm{inc}=250\,\mathrm{keV}$. The values have been calculated by subtracting the mean energy of the scattered spectra from the incident energy. All these values have been obtained by Gaussian fits to the main peak of the spectra. The individual data points for each incidence angle are connected by straight lines.}
	\label{fig:plot250keVeloss}
\end{figure}

\begin{figure}
  \includegraphics[width=.8\textwidth]{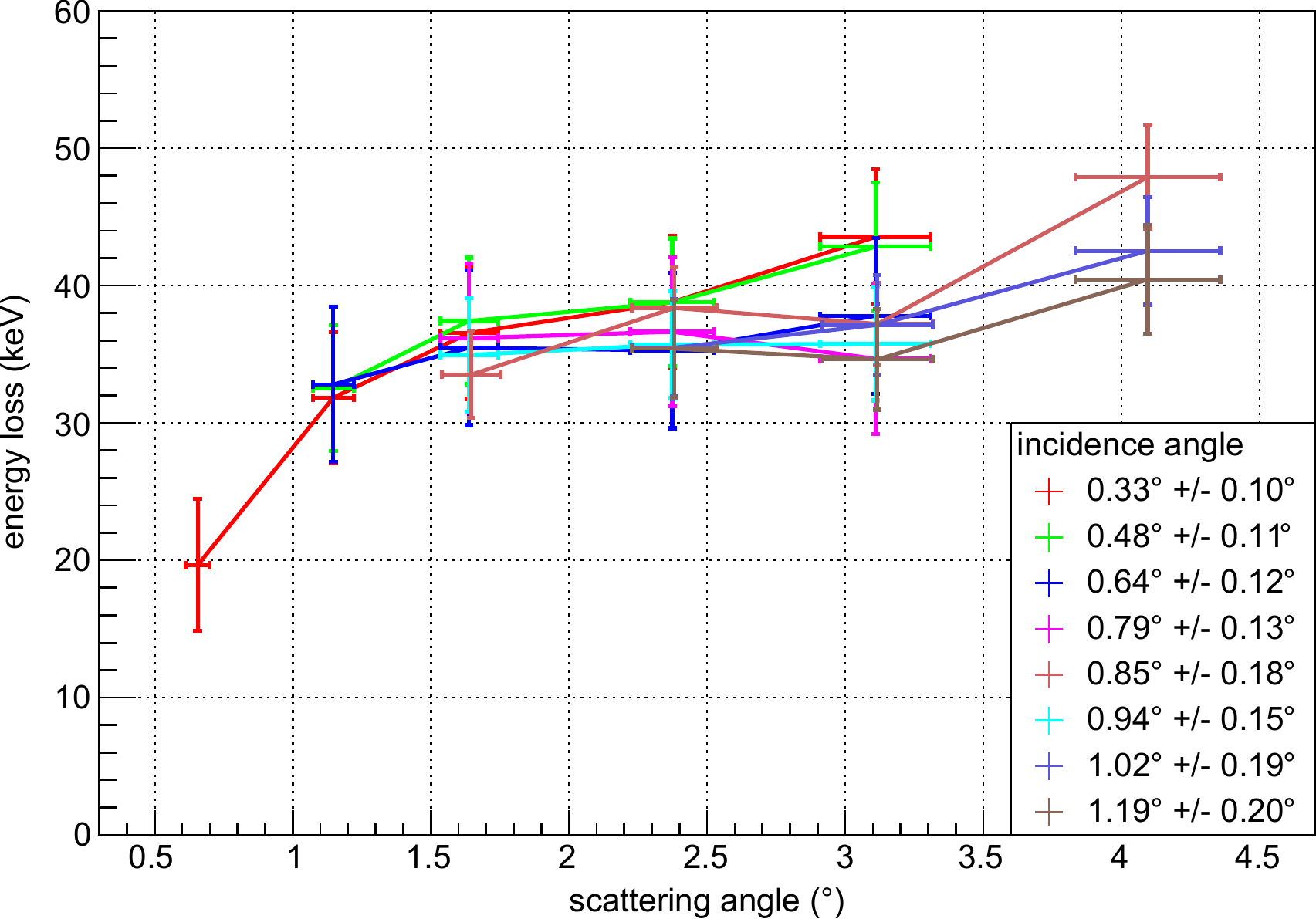}
	\caption{Same as Fig.~\ref{fig:plot250keVeloss}, but for $E_\mathrm{inc}=500\,\mathrm{keV}$.}
	\label{fig:plot500keVeloss}
\end{figure}

\begin{figure}
  \includegraphics[width=.8\textwidth]{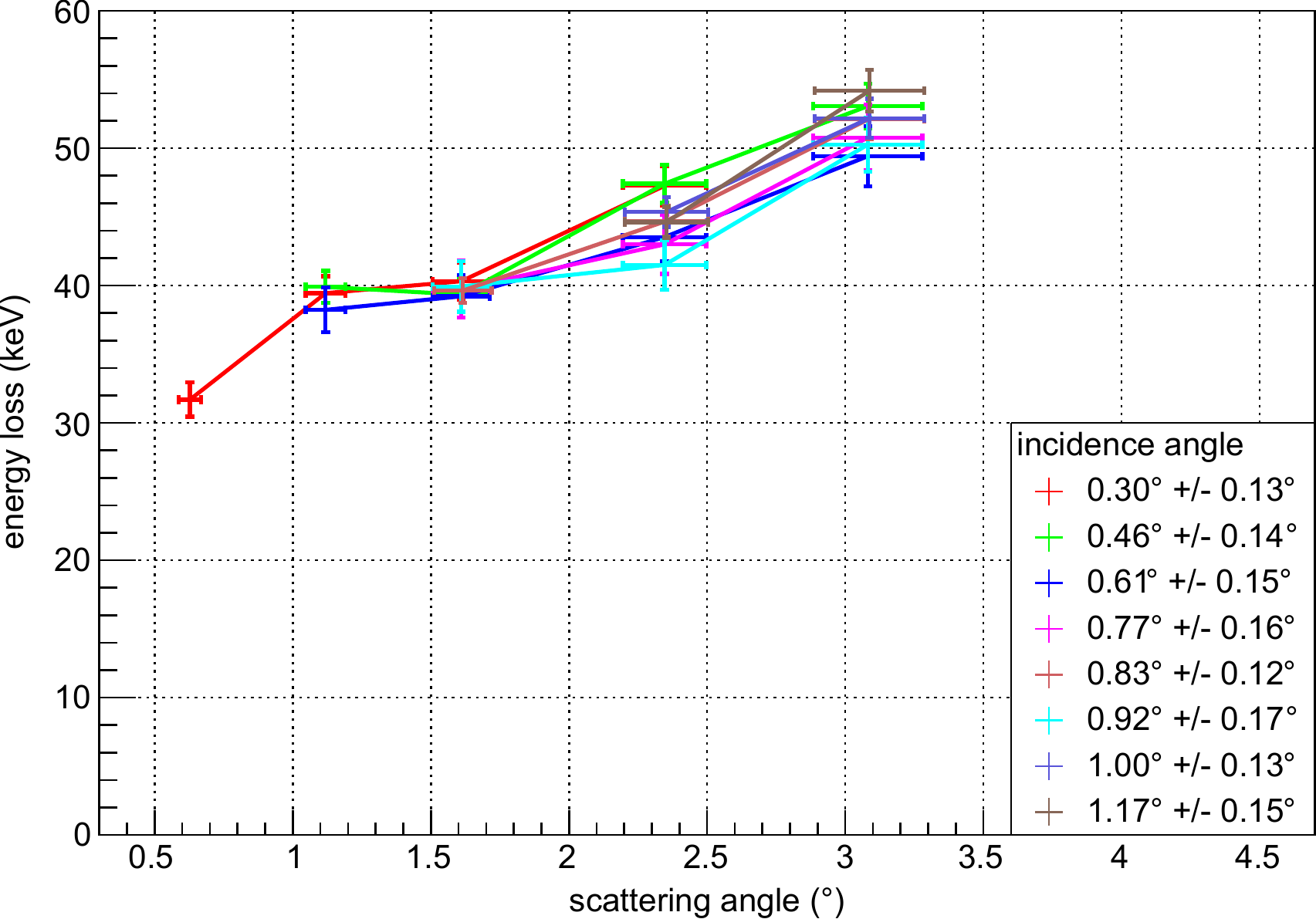}
	\caption{Same as Fig.~\ref{fig:plot250keVeloss}, but for $E_\mathrm{inc}=1\,\mathrm{MeV}$.}
	\label{fig:plot1MeVeloss}
\end{figure}

The results of the energy loss analysis is presented in Figs.~\ref{fig:plot250keVeloss}\textendash\ref{fig:plot1MeVeloss}. The plots show the most probable energy loss $\Delta E$, which is derived from Gaussian fits of the main peaks of the incident and the scattered spectra. Since shifts of the beam energy of a few keV can occur during the measurements, the incident spectrum has been measured immediately before and after each set of scattering measurements. The mean of these has been used as incident energy, from which the peak energy of the scattered spectra has been subtracted. For the calculation of the errors the deviation of the energy before and after the measurement as well as the errors of the fits have been considered. The former is dominating the errors, especially in the measurement around 250\,keV. The angular errors have been derived as described in Section~\ref{ssec:eff}.

Since the Firsov scattering description makes no assumption of the energy loss, a narrow Gaussian distribution around 3\,keV, independent of the incident energy has been used as default value in the implementation of the Firsov process in \textsc{Geant4}. This assumption originates from high precision experimental data of grazing incidence proton scattering on a planar aluminum surface \cite{Winter_1997}. However, the present results for the gold coated \textit{eROSITA} mirrors indicate significant higher losses: from 13\,keV for $\Theta = 0.69$\textdegree~and $E_{\mathrm{inc}} = 250\,\mathrm{keV}$ up to 54\,keV for $\Theta = 3.09$\textdegree~and 1\,MeV. It can be concluded that the most probable energy loss depends on the incident energy, similar to the results obtained with graphite targets reported in \cite{Pfandzelter_1992}. Furthermore, the data show a significant increase of the energy loss towards larger scattering angles, while being independent of the incident angle. An explanation for the discrepancy among the experimental results is first of all the different target materials as well as different surface conditions of the mirror samples used in the present study compared to the extensively prepared graphite and aluminum targets, which feature a clearly defined crystal orientation and flatness on an atomic level without contamination from adsorbed gases.

\subsection{Comparison to the Firsov description and simulations}\label{ssec:comparison}
The theoretical scattering description by Firsov \cite{Firsov} (cf. Eq.~\ref{eq:Firsov})\footnote{The scattering angle has been transformed to match the convention used in this study.}, which has been implemented in \textsc{Geant4} \cite{Lei_2004}, predicts scattering distributions for grazing angle proton scattering that are shifted to smaller scattering angles in comparison to the \textsc{Geant4} multiple scattering process; the Firsov distributions peak at scattering angles that meet the condition for specular reflection ($\Theta = 2\Psi$). In order to compare the experimental data of the present study to the Firsov formula, the integral of the measured distributions from $\Theta = \Psi$, where zero efficiency has been assumed, to the largest scattering angle in the measurement has been estimated by a linear interpolation between the data points. The dashed curves in Figs.~\ref{fig:plot250keVeff}\textendash\ref{fig:plot1MeVeff} have been calculated by normalizing the integral of the Firsov curve for the particular incidence angle to the integral of the respective measurement, thereby equalizing the integral scattering efficiency in the considered range of scattering angles. The agreement with the experimental data is reasonable, considering that the mirror surface is not ideal, the uncertainties of the measurement, and the linear interpolation used for the comparison. However, the measured peak efficiencies are slightly shifted to larger scattering angles and the slope for larger scattering angles is steeper in the data than predicted by the formula. This should be taken into account in simulations of the proton propagation through an X-ray telescope, where two scatters can occur.

\begin{equation}
\epsilon(\Psi, \Theta) = \frac{3(\Psi(\Theta-\Psi))^{3/2}}{2\pi\Psi(\Psi^3+(\Theta-\Psi)^3)}
\label{eq:Firsov}
\end{equation}

\begin{figure}
  \includegraphics[width=.8\textwidth]{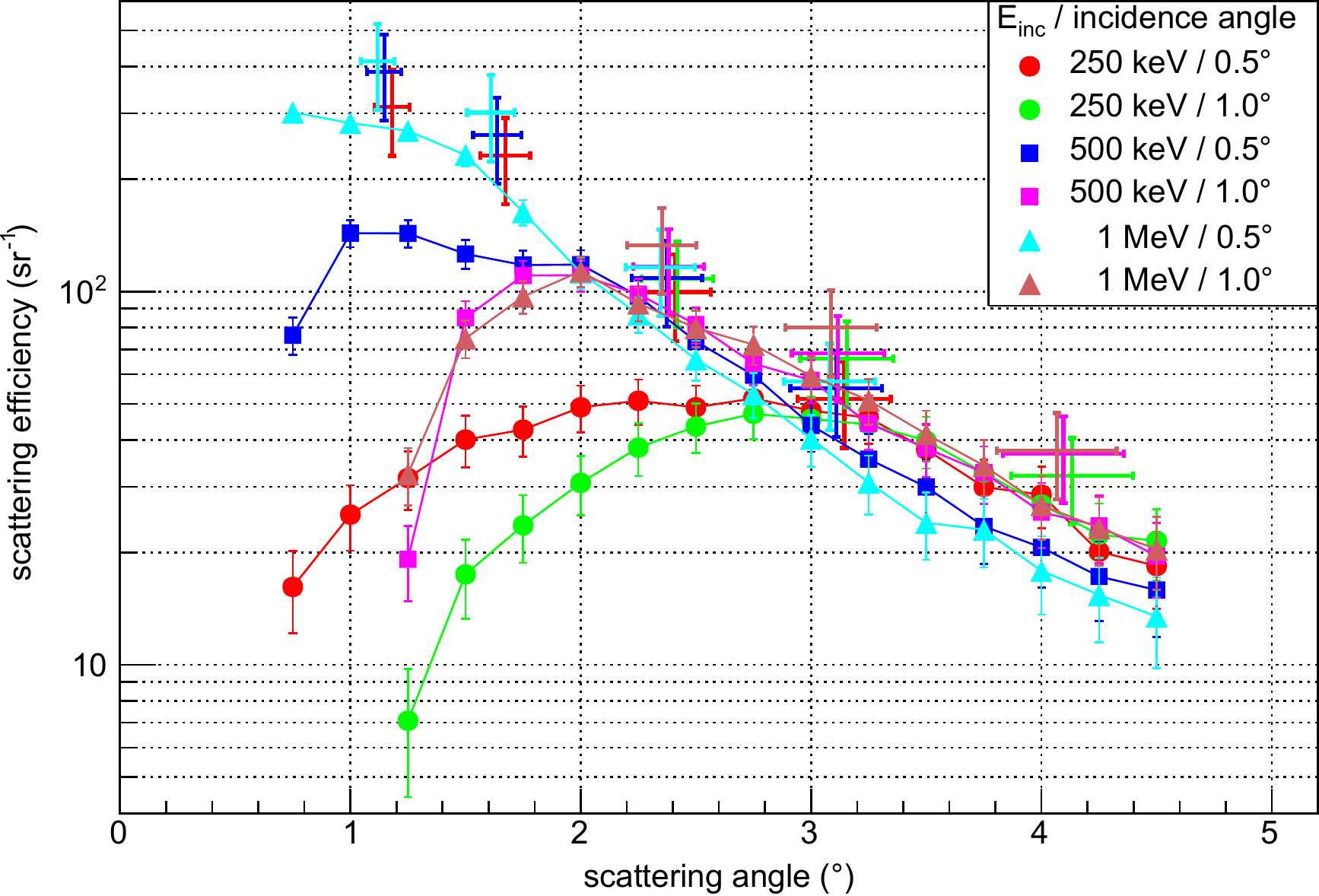}
	\caption{Scattering efficiency results of \textsc{TRIM} simulations, each with 10 million protons. Angular cuts and an energy cut below 100\,keV have been applied to match the acceptance angle and the energy threshold of the detector in the experimental setup. The individual data points of one parameter set are connected by straight lines. Included are the experimental data points of the measurement sets that correspond best to the parameters of the simulations (drawn in the respective color).}
	\label{fig:TRIM_eff}
\end{figure}

\begin{figure}
  \includegraphics[width=.8\textwidth]{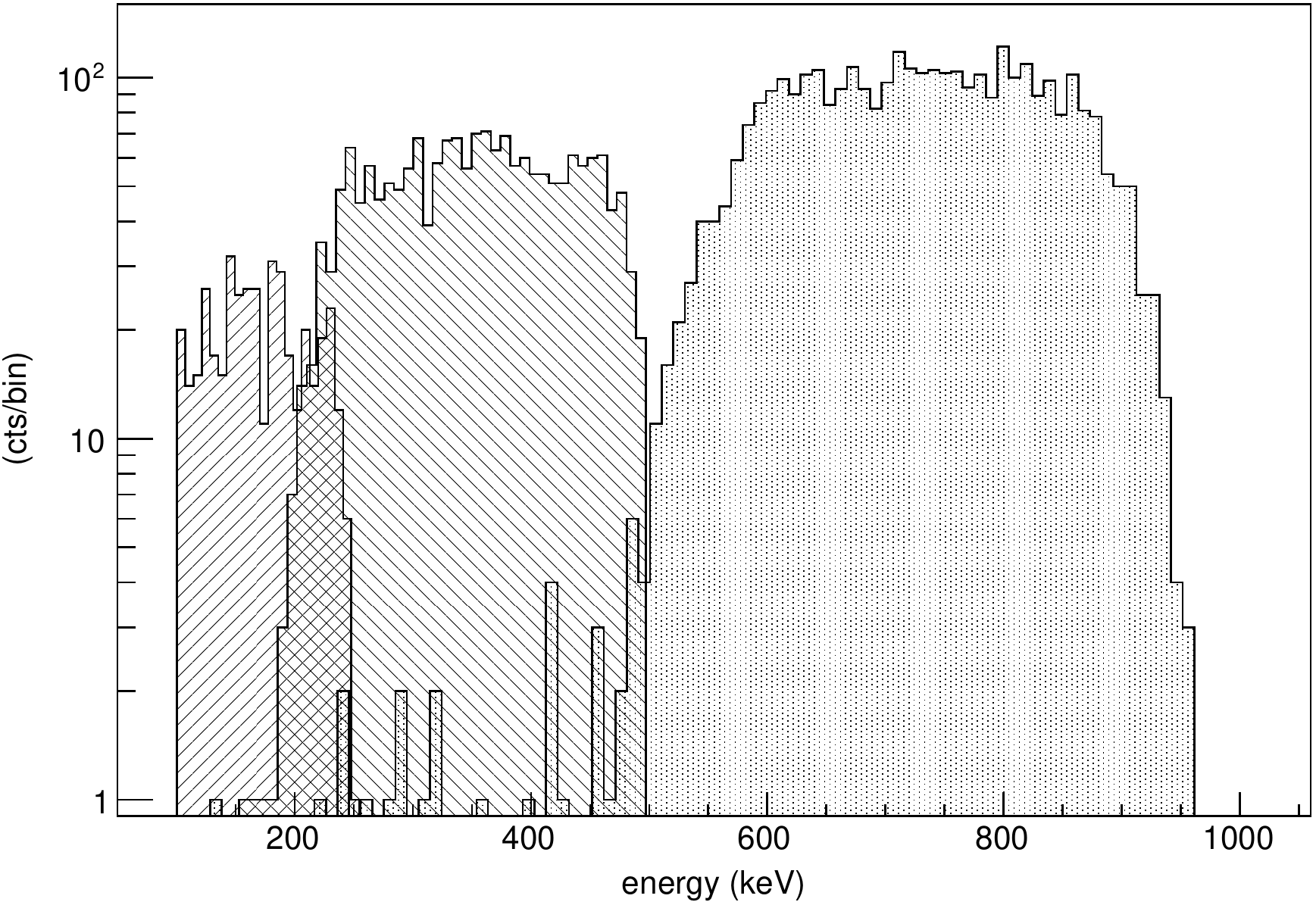}
	\caption{Proton spectra after scattering obtained by means of \textsc{TRIM} simulations. From left to right the incidence energy is 250\,keV, 500\,keV, and 1\,MeV, respectively. The incidence angle is 0.5\textdegree, the scattering angle 2\textdegree. Angular cuts and an energy cut below 100\,keV have been applied to match the acceptance angle and the energy threshold of the detector in the experimental setup.}
	\label{fig:TRIM_eloss_.5}
\end{figure}

\begin{figure}
  \includegraphics[width=.8\textwidth]{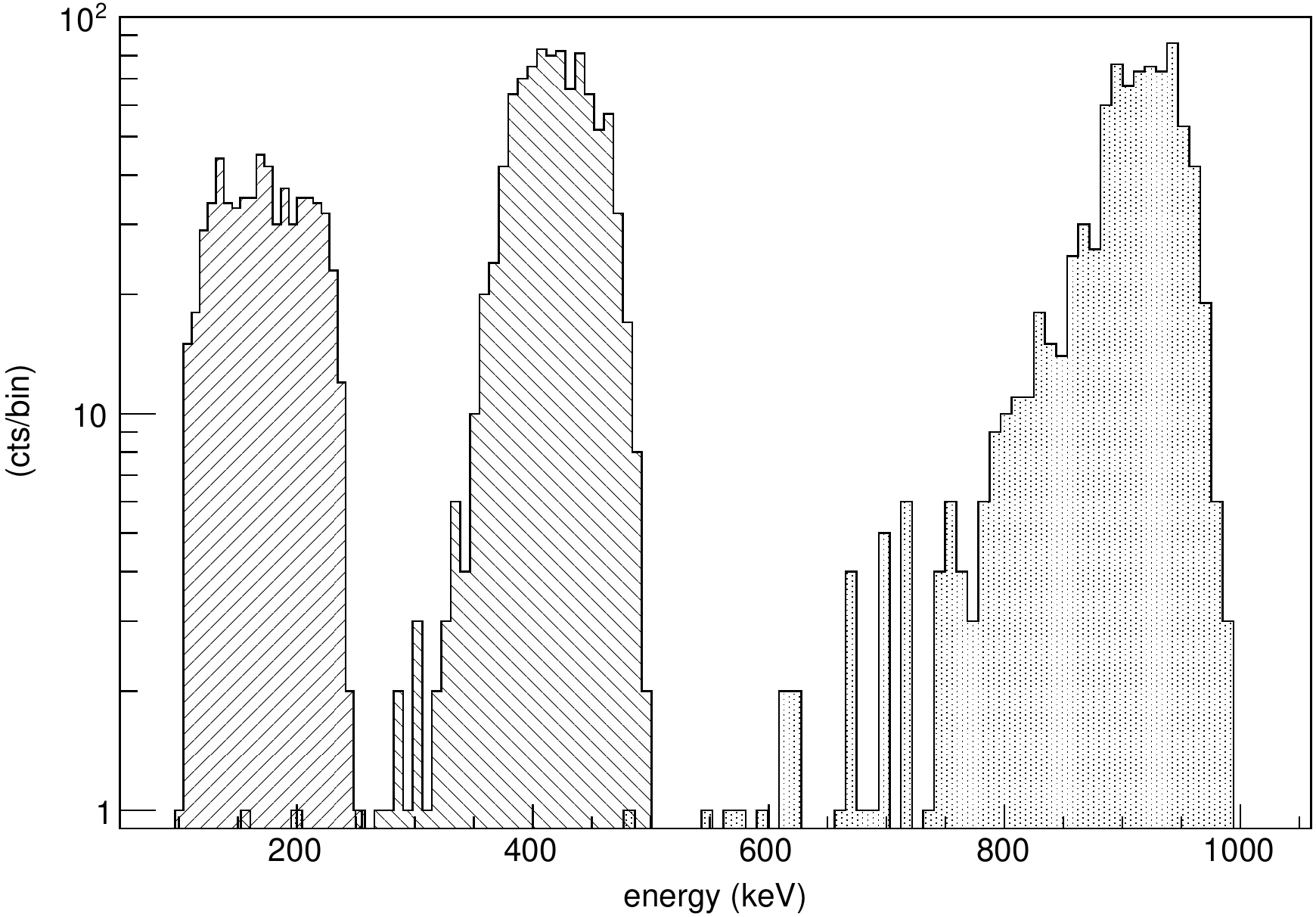}
	\caption{Same as Fig.~\ref{fig:TRIM_eloss_.5}, but for $\Psi=1.0$\textdegree.}
	\label{fig:TRIM_eloss_1}
\end{figure}

The data have been compared to \textsc{TRIM} simulations. Therefore, the scattering of 10 million protons on a flat target, consisting of a 50\,nm gold layer on a nickel substrate, has been simulated for each of the three incident energies by using incidence angles of 0.5\textdegree~and 1.0\textdegree, respectively. Angular cuts and an energy cut at 100\,keV have been applied to the simulation output to match the acceptance angle and the energy threshold of the detector in the experimental setup. The results for the scattering efficiency are shown in Fig.~\ref{fig:TRIM_eff}. \textsc{TRIM} reproduces the experimental data for the larger scattering angles, except for a slight systematic shift to lower efficiencies. Nevertheless, below a certain scattering angle the simulated efficiencies drop significantly with respect to the experimental data. For 500\,keV incident energy the threshold is around a scattering angle of 2\textdegree, for 250\,keV around 3\textdegree. Just the simulation for 1\,MeV and an incidence angle of 0.5\textdegree~is in agreement with the data down to a scattering angle of 1\textdegree. Furthermore, the \textsc{TRIM} output shows significantly larger energy losses and broader distributions than the measured data. As shown in Figs.~\ref{fig:TRIM_eloss_.5} and \ref{fig:TRIM_eloss_1}, this is more pronounced for smaller incidence angles. The broad energy loss distribution accounts partially for the low efficiencies at 250\,keV, since a part of the scattered spectrum is affected by the 100\,keV energy cut.

Since the compositions of the \textit{eROSITA} and the \textit{XMM-Newton} mirrors are almost identical, the measured efficiencies have also been compared to the \textsc{Geant4} results reported in \cite{Chandra_XMM}. In general, the simulation results show smaller efficiencies than the current experiment; deviations up to a factor of two occur in particular for large incidence angles, while for smaller incidence and larger scattering angles the results agree within the errors. Even though this behavior appears to be similar to the \textsc{TRIM} results discussed before, the overlap in parameter space is too small to allow significant conclusions.

\section{Conclusions}\label{sec:conc}
The scattering setup presented in this work applies a unique method of determining the incident proton flux on the target via monitor detectors, enabling high accuracy efficiency measurements of grazing incidence scattering. The current data obtained with samples of \textit{eROSITA} mirror shells show significant higher efficiencies compared to previous measurements with \textit{XMM-Newton} mirrors at the Harvard University in 1999 \cite{XMM-RGS}. An explanation, at least for a part of the discrepancy, could be the different methods for the determination of the incident flux.

The use of an energy degrader foil, which leads to a spectral broadening, and the moderate energy resolution of the SSB detectors limit the precision of energy loss measurements with the present setup. However, since the measured mean energy loss is more than one order of magnitude larger than estimated in \cite{Lei_2004}, the deviation is significant.

It can be concluded that the scattering distributions of the present study are reasonably reproduced by the Firsov formula. However, a small systematic shift of the measured peak efficiencies to larger scattering angles and a steeper slope of the experimental data has been found. These deviations might be attributed to the non-ideal finite surface roughness of the mirror shells and a combination of Firsov scattering with other scattering processes, e.g. multiple scattering, might yield a more accurate description. If incident energies and scattering angles are sufficiently large, simulations by means of the \textsc{TRIM} code are in good agreement with the measured data. The lower thresholds for the applicability of \textsc{TRIM} for efficiency simulations are around a scattering angle of 3\textdegree~for 250\,keV incident energy and at about 2\textdegree~for 500\,keV and 1\,MeV. For 1\,MeV incident energy and an incidence angle of 0.5\textdegree~the agreement between simulation and measurement extends even down to scattering angles of 1\textdegree. However, for all energies and angles studied within this work the energy loss in the simulation is larger and broader distributed than in the experimental results. As for the efficiencies, the deviation is decreasing with increasing energy and angles.

Future applications of the setup are grazing angle scattering measurements with different target materials that are used for X-ray mirror coatings as well, e.g. iridium, and aluminum for a comparison with a light target material. Additionally, measurements with alpha particles are planned. The results of this work will be incorporated in a ray tracing simulation code for protons that is currently under development in a collaboration of the IASF\footnote{Istituto di Astrofisica Spaziale e Fisica Cosmica} in Palermo, the IAA\footnote{Institute for Astronomy and Astrophysics} T\"ubingen, and the MPE\footnote{Max Planck Institute for Extraterrestrial Physics} in Garching \cite{Perinati_2014}. It is dedicated to the simulation of proton propagation through X-ray telescopes for the assessment of potential damage and proton induced background in future X-ray missions, such as \textit{eROSITA} and \textit{ATHENA}.

\begin{acknowledgements}
This work is partially supported by the Bun\-des\-mi\-ni\-ste\-ri\-um f\"ur Wirt\-schaft und Technologie through the Deutsches Zentrum f\"ur Luft- und Raumfahrt (Grant FKZ 50 OO 1110).
\end{acknowledgements}
\small{\textbf{Conflict of Interest} The authors declare that they have no conflict of interest.}



\end{document}